\documentclass[twocolumn]{aastex63}
\usepackage{array,multirow,graphicx}
\usepackage{float}
\usepackage{amsmath}

\newcommand{\Var}{{\mathbb V}}

\newcommand{\Cov}{\mathrm{Cov}}

\newcommand{\nsncand}{3088 }
\newcommand{\nspec}{408 }
\newcommand{\nsnany}{401 }
\newcommand{\nsnia}{268 }
\newcommand{\nsnii}{86 }
\newcommand{\nsnibc}{29 }
\newcommand{\nsnpec}{16 }
\newcommand{\nsnnoclass}{two }
\newcommand{\nunclear}{two }
\newcommand{\ngalsn}{one }

\newcommand{\nnosn}{four }
\newcommand{\nsnrisegtptfour}{64 }
\newcommand{\nsnrisegtone}{ten }
\newcommand{\nLCclf}{$8.5\times10^5$ }
\newcommand{\nstampclf}{$1.9\times10^7$ }
\newcommand{\nobjects}{$3.7\times10^7$ }
\newcommand{\nalerts}{$9.7\times10^7$ }
\newcommand{\nnondet}{$1.1\times10^9$ }
\newcommand{\nusers}{2788 }
\newcommand{\ncountries}{52 }

\shorttitle{The ALeRCE broker}
\shortauthors{F\"orster et al.}
\graphicspath{{./}{figures/}}

\begin{document}

\title{The Automatic Learning for the Rapid Classification of Events (ALeRCE) Alert Broker}

\newcommand\MAS{Millennium Institute of Astrophysics, Nuncio Monse{\~{n}}or S{\'{o}}tero Sanz 100, Providencia, Santiago, Chile}
\newcommand\CMM{Center for Mathematical Modeling, University of Chile, AFB170001, Chile}
\newcommand\ASTROUCH{Departamento de Astronom\'ia, Universidad de Chile, Casilla 36D, Santiago, Chile}
\newcommand\ASTROUC{Instituto de Astrof{\'{\i}}sica, Facultad de F{\'{i}}sica, Pontificia Universidad Cat{\'{o}}lica de Chile, Av. Vicu\~{n}a Mackenna 4860, 7820436 Macul, Santiago, Chile} 
\newcommand\AIUCUC{Centro de Astroingenier{\'{\i}}a, Pontificia Universidad Cat{\'{o}}lica de Chile, Av. Vicu\~{n}a Mackenna 4860, 7820436 Macul, Santiago, Chile}
\newcommand\DIEUCH{Department of Electrical Engineering, Universidad de Chile, Av. Tupper 2007, Santiago 8320000, Chile}
\newcommand\CSUDEC{Department of Computer Science, University of Concepción, Chile}
\newcommand\UAI{Department of Engineering and Science, Universidad Adolfo Ibañez, Av. Diagonal Las Torres 2700, Santiago, Chile}
\newcommand\ASTROUV{Instituto de F\'isica y Astronom\'ia, Universidad de Valpara\'iso, Av. Gran Breta\~na 1111, Playa Ancha, Casilla 5030, Chile}
\newcommand\CSUTEM{Departmento de Inform{\'{a}}tica y Computaci{\'{o}}n, Universidad Tecnol\'{o}gica Metropolitana, 
Santiago, Chile}
\newcommand\IOA{Institute of Astronomy, University of Cambridge, Madingley Road, Cambridge CB3 0HA, UK}
\newcommand\Caltech{Division of Physics, Mathematics, and Astronomy, California Institute of Technology, Pasadena, CA 91125, USA}
\newcommand\IACS{Institute for Applied Computational Science, Harvard University, Cambridge, MA, USA}
\newcommand\CDDD{Center for Data Driven Discovery, California Institute of Technology, Pasadena, CA 91125, USA}

\correspondingauthor{F.~F\"orster}
\email{francisco.forster@gmail.com}

\collaboration{100}{(The ALeRCE collaboration)}

\author[0000-0003-3459-2270]{F.~F\"orster}
\affil{\CMM}
\affil{\MAS}
\affil{\ASTROUCH}

\author [0000-0002-2720-7218]{G.~Cabrera-Vives}
\affil{\CSUDEC}
\affil{\MAS}

\author[0000-0001-9164-4722]{E.~Castillo-Navarrete}
\affil{\CMM}
\affil{\MAS}

\author[0000-0001-9164-4722]{P.~A.~Est\'evez}
\affiliation{\DIEUCH}
\affil{\MAS}
\affil{\CMM}

\author[0000-0003-0820-4692]{P.~S\'anchez-S\'aez}
\affil{\MAS}
\affil{\ASTROUC}
\affil{\UAI}

\author[0000-0002-2045-7134]{J.~Arredondo}
\affil{\MAS}

\author[0000-0002-8686-8737]{F.~E.~Bauer} 
\affil{\ASTROUC},\affil{\AIUCUC}, \affil{\MAS}, \affil{Space Science Institute, 4750 Walnut Street, Suite 205, Boulder, Colorado 80301, USA}

\author[0000-0003-4673-8791]{R.~Carrasco-Davis}
\affil{\MAS}
\affiliation{\DIEUCH}

\author[0000-0001-6003-8877]{M.~Catelan}
\affil{\ASTROUC}
\affil{\AIUCUC}
\affil{\MAS}

\author[0000-0002-1835-7433]{F.~Elorrieta}
\affil{\MAS}
\affil{Department of Mathematics and Computer Science, Universidad de Santiago de Chile, Av. Libertador Bernardo O’Higgins 3663, Estaci\'on Central, Santiago, Chile}

\author[0000-0003-4723-9660]{S.~Eyheramendy}
\affil{\MAS}
\affiliation{\UAI}

\author[0000-0003-3541-1697]{P.~Huijse}
\affiliation{Instituto de Inform\'atica, Universidad Austral de Chile, General Lagos 2086, Valdivia, Chile}
\affil{\MAS}

\author[0000-0001-6003-8877]{G.~Pignata}
\affil{Departamento de Ciencias F\'isicas, Universidad Andres Bello, Avda. Republica 252, Santiago, Chile}
\affil{\MAS}

\author[0000-0003-3455-9358]{E.~Reyes}
\affil{\MAS}
\affiliation{\DIEUCH}

\author[0000-0003-3627-0216]{I.~Reyes}
\affil{\MAS}
\affil{\CMM}
\affiliation{\DIEUCH}

\author{D.~Rodr\'iguez-Mancini}
\affil{\MAS}
\affil{\CSUDEC}

\author[0000-0002-1292-2374]{D.~Ruz-Mieres}
\affil{\CMM}
\affil{\MAS}
\affil{\IOA}

\author [0000-0001-5306-1390]{C.~Valenzuela}
\affil{\MAS}
\affil{\CMM}

\author {I.~\'Alvarez-Maldonado}
\affil{\MAS}
\affil{\CMM}

\author {N.~Astorga}
\affil{\MAS}
\affiliation{\DIEUCH}

\author [0000-0002-5936-7718]{J.~Borissova}
\affil{\ASTROUV}
\affil{\MAS}

\author [0000-0002-5936-7718]{A.~Clocchiatti}
\affil{\ASTROUC}
\affil{\MAS}

\author[0000-0001-7208-5101]{D.~De Cicco}
\affil{\MAS}
\affil{\ASTROUC} 

\author{C.~Donoso-Oliva}
\affil{\CSUDEC}
\affil{\MAS}

\author[0000-0002-3168-0139]{M.~J.~Graham}
\affil{\Caltech}

\author [0000-0002-9740-9974]{R.~Kurtev}
\affil{\ASTROUV}
\affil{\MAS}

\author[0000-0003-2242-0244]{A.~Mahabal}
\affiliation{\Caltech}
\affiliation{\CDDD}

\author{J.C. Maureira}
\affiliation{\CMM}

\author{R.~Molina-Ferreiro}
\affiliation{\DIEUCH}

\author{A.~Moya}
\affiliation{\MAS}
\affiliation{\CMM}

\author{W.~Palma}
\affiliation{\MAS}

\author{M.~Pérez-Carrasco}
\affil{\CSUDEC}
\affil{\MAS}

\author[0000-0002-8178-8463]{P.~Protopapas}
\affil{\IACS}

\author {M.~Romero}
\affiliation{\DIEUCH}

\author [0000-0002-1614-9825]{L.~Sabatini-Gacitua}
\affil{\MAS}
\affil{\CMM}

\author {A.~Sánchez}
\affil{\CSUDEC}
\affil{\MAS}

\author{J.~San Mart\'in}
\affil{\CMM}

\author{C.~Sep\'ulveda-Cobo}
\affil{\MAS}
\affil{\CMM}

\author{E.~Vera}
\affil{\CMM}

\author[0000-0001-6699-4181]{J.~R.~Vergara}
\affil{\CSUTEM}
\affil{\MAS}



\begin{abstract}

We introduce the Automatic Learning for the Rapid Classification of Events (ALeRCE) broker, an astronomical alert broker designed to provide a rapid and self--consistent classification of large etendue telescope alert streams, such as that provided by the Zwicky Transient Facility (ZTF) and, in the future, the Vera C. Rubin Observatory Legacy Survey of Space and Time (LSST). ALeRCE is a Chilean--led broker run by an interdisciplinary team of astronomers and engineers, working to become intermediaries between survey and follow--up facilities. ALeRCE uses a pipeline which includes the real--time ingestion, aggregation, cross--matching, machine learning (ML) classification, and visualization of the ZTF alert stream. We use two classifiers: a stamp--based classifier, designed for rapid classification, and a light--curve--based classifier, which uses the multi--band flux evolution to achieve a more refined classification. We describe in detail our pipeline, data products, tools and services, which are made public for the community (see \url{https://alerce.science}). Since we began operating our real--time ML classification of the ZTF alert stream in early 2019, we have grown a large community of active users around the globe. We describe our results to date, including the real--time processing of \nalerts alerts, the stamp classification of \nstampclf objects, the light curve classification of \nLCclf objects, the report of \nsncand supernova candidates, and different experiments using LSST-like alert streams. Finally, we discuss the challenges ahead to go from a single-stream of alerts such as ZTF to a multi--stream ecosystem dominated by LSST.

\end{abstract}




\section{Introduction}


The exponential growth of the light collecting area of telescopes and the number of pixels of digital detectors has resulted in a new generation of survey telescopes that are revolutionizing the way we study the time domain in astronomy \citep{Anthony_Tyson_2019}. New surveys that systematically scan the optical/near infrared sky with deep, wide and fast cadence observations (e.g., Catalina Real-Time Transient Survey, CRTS, \citealt{2009ApJ...696..870D}; Palomar Transient Factory, PTF, \citealt{2009PASP..121.1395L}; Optical Gravitational Lensing Experiment, OGLE, \citealt{2015AcA....65....1U}; Dark Energy Survey, DES, \citealt{2005astro.ph.10346T}; SkyMapper, \citealt{2007PASA...24....1K}; {\em Kepler}, \citealt{2010ApJ...713L..79K}; Vista Variables in the Via Lactea Survey, VVV, \citealt{2010NewA...15..433M}; Korea Microlensing Telescope Network, KMTNet, \citealt{2016JKAS...49...37K}; Hyper Suprime-Cam Subaru Strategic Program, HSC-SSP, \citealt{10.1093/pasj/psx066}; Asteroid Terrestrial--Impact Last Alert System, ATLAS, \citealt{2018PASP..130f4505T}; Zwicky Transient Facility, ZTF, \citealt{2019PASP..131a8002B}; Deeper, Wider, Faster, DWF, \citealt{2020MNRAS.491.5852A}) are uncovering large populations of time--varying astrophysical phenomena, including new populations of dim, rare, and/or short-lived events \citep[e.g.,][]{2012ApJ...755..161K, 2014ApJ...794...23D}. 

Meanwhile, the construction of the Vera C. Rubin Observatory and its Legacy Survey of Space and Time, LSST \citep{2009arXiv0912.0201L}, is advancing, and a convergence is expected to happen with surveys in other regions of the electromagnetic spectrum  (e.g., Square Kilometer Array, SKA,  \citealt{2009IEEEP..97.1482D}; Wide-field Infrared Survey Explorer, WISE, \citealt{2010AJ....140.1868W}; eROSITA, \citealt{2012arXiv1209.3114M}; Fermi Gamma-ray Space Telescope, \citealt{2009ApJ...697.1071A};  Cherenkov Telescope Array, CTA, \citealt{2011ExA....32..193A}), high energy particles (e.g., CTA; IceCube Neutrino Observatory, \citealt{2017JInst..12P3012A}), and gravitational waves (Laser Interferometer Gravitational-Wave Observatory, \citealt{1992Sci...256..325A}; Advanced Virgo, \citealt{2015CQGra..32b4001A}), opening a new era of multi--messenger astronomy \citep{2017ApJ...848L..12A,2018Sci...361..147I}.

\begin{figure*} [htb!]
\centering
\includegraphics[width=1.0\textwidth]{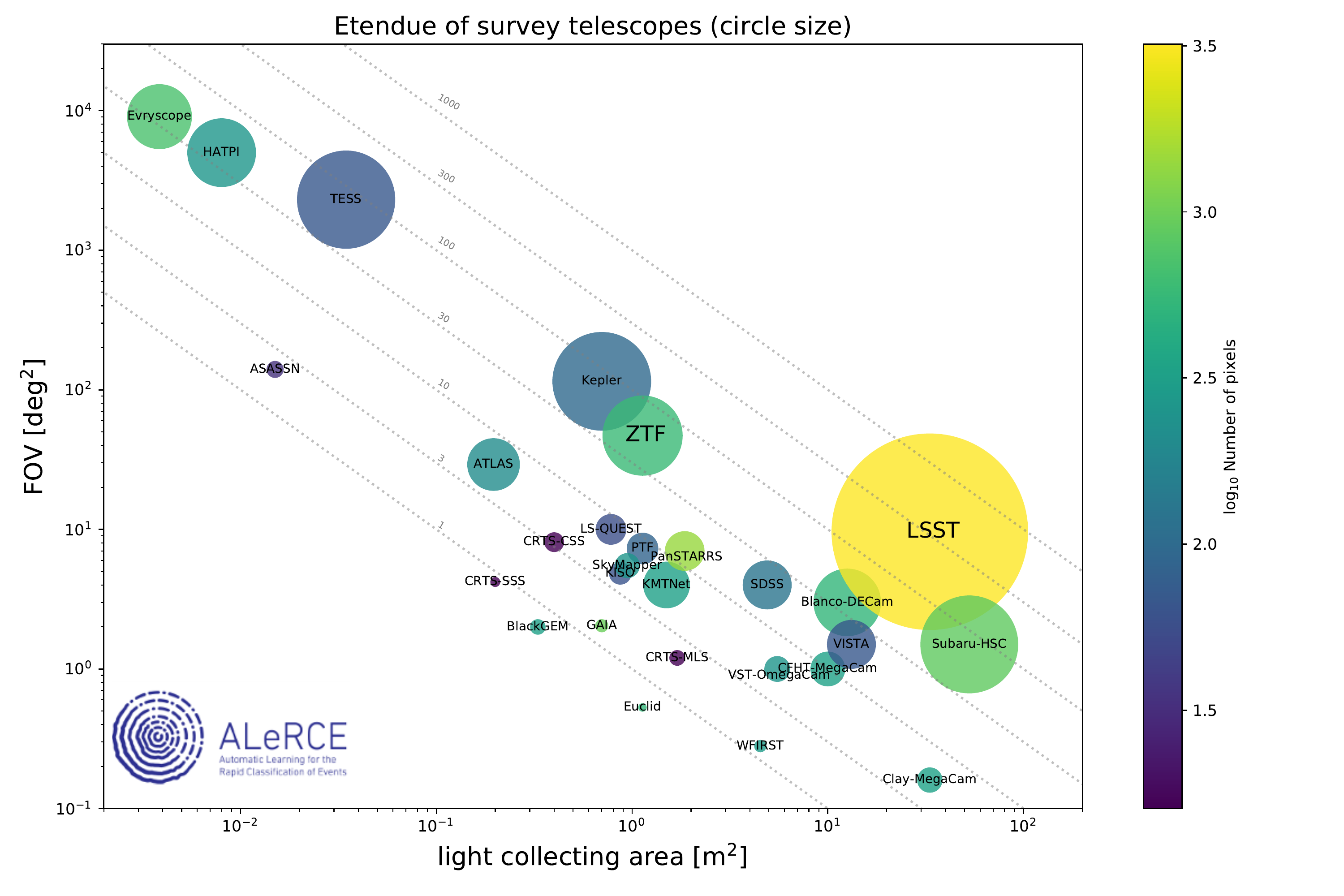}
\caption{\label{fig:etendue} FOV vs light collecting area for a selection of ground and space-based survey telescopes currently operational or planned. The product of the two is called etendue and is indicated by the relative sizes. Note that if a survey contains several identical telescopes we consider the sum of their etendues. The color of the circles indicates the number of pixels in the main camera of the instrument, following the color coding on the right. Constant etendue loci are shown as gray dashed lines, with the specific etendue value shown for each line. See Table~\ref{tab:etendue} for telescope names and references.}
\end{figure*}

The fundamental quantity that defines a survey telescope is the product of mirror area and field of view (FOV), known as etendue, which is a simple proxy for the volume in space that can be monitored by different telescopes for the same exposure time and for a given intrinsic lumininosity object. We show the FOV, collecting area and number of pixels of a selection of large etendue survey telescopes in Figure~\ref{fig:etendue}. These telescopes vary from very large FOV or all--sky collections of small aperture telescopes (``hedgehog" configurations) to large aperture and large FOV detector mosaics (e.g., LSST). The small aperture telescopes are able to explore very fast cadences, but are restricted in practice to bright objects or the nearby Universe. The large aperture telescopes are able to explore dimmer objects and the more distant Universe, but have more restricted cadences for all--sky observations.

The detectors in these large etendue telescopes produce data at increasingly faster rates. Millions of \textit{events}, i.e., objects that are witnessed to change their brightness or position in the sky, are being detected and reported in the form of continuous astronomical alert streams \citep{2019PASP..131a8001P}. These streams create an opportunity for a new generation of follow--up telescopes to characterize large numbers of astronomical events in a coordinated fashion, ultimately leading to a better understanding of the nature of variable phenomena and consequently of the evolution of our local and more distant Universe. 

A new time--domain ecosystem is being built accordingly, where telescopes specialize as either survey or follow--up telescopes, but also where new digital information components are developed to connect them seamlessly. The aggregation, annotation and classification of alerts in a rapid and consistent fashion is done by  \emph{astronomical alert brokers}, such as the Automatic Learning for the Rapid Classification of Events, ALeRCE, this work; Alert Management, Photometry and Evaluation of Lightcurves, AMPEL \citep{2019A&A...631A.147N}; Arizona-NOAO Temporal Analysis and Response to Events System, ANTARES \citep{2018ApJS..236....9N}; Fink;\footnote{\url{https://fink-broker.readthedocs.io/}} LASAIR;\citep{2019RNAAS...3...26S} and Make Alerts Really Simple, MARS.\footnote{ \url{https://mars.lco.global/}}  Different brokers typically specialize in different science cases. Their main role is to provide a fast and consistent classification of the alert stream using all the available data, but also to enable filtering of the stream for different scientific communities. The fast classification of events is critical for the study of either short--lived phenomena or the early phases of evolution of longer--lived processes, enabling follow--up observations to occur fast enough for some physical properties to be inferred \citep[e.g.,][]{2014Natur.509..471G}. They will also contribute to the detection of new astrophysical phenomena in the form of outliers/anomalies \citep[e.g.,][]{2016AJ....152...71N}, and will help reveal new sub-populations among known families of events \citep[e.g.,][]{2017MNRAS.465.4530B}. 

An interoperable and agile ecosystem is needed, with all the relevant parts able to interact automatically to perform coordinated observations, but also capable of adapting quickly to new science cases, instruments, or digital technologies. In this new scenario, follow--up telescopes will listen and react to Target and Observation Managers \citep[TOMs; e.g.,][]{2018SPIE10707E..11S}. TOMs will listen to alert broker classified streams, and brokers will listen to survey telescope alert streams. When follow-up observations are performed and their results become available, TOMs will be able to modify their follow-up strategy, brokers will be able to improve their classification, and survey telescopes will be able to change their surveying strategies, providing a feedback mechanism for the entire time domain ecosystem to continuously improve.

\subsection{Alert Broker Challenges}

Astronomical alert brokers are a new kind of tool in the interface between astronomy and data science. They face new challenges including infrastructure, machine learning (ML), and community integration, but also organizational aspects which are important in order to effectively add value to the community. This makes them important laboratories for testing new ideas on data science going even beyond astronomy.

In terms of infrastructure, the biggest challenge for astronomical brokers is to ingest, annotate and classify, in a scalable fashion, the large astronomical alert streams coming from large etendue telescopes such as ZTF or LSST. For example, we have received typically between 10$^5$--10$^6$ alerts per night from the public ZTF stream, associated with \nobjects objects as of Jun 2020. For comparison, LSST is expected to produce about 10$^7$ alerts per night and contain more than 10$^9$ different objects, which requires a distributed type of database and processing. Additionally, there will be a diversity of surveys streaming alerts which must be cross--matched and classified in real time (e.g., ZTF, ATLAS, LSST). Thus, the challenge is to ingest data streams from a diversity of telescopes in a scalable fashion and to classify them using their combined information to enable a rapid reaction by follow--up telescopes and a self--consistent analysis.

In terms of ML development, the challenges are diverse. What is an appropriate and relevant taxonomy for the astronomical community? How should we balance classification purity and efficiency? How can we develop ML classifiers and bring them into production in a reasonable timescale? How should we include cross--matched information in these classifiers? How can we train models using data which may be highly unbalanced and not fully representative of the unlabeled data? For example, training a classifier with spectroscopically labeled data will tend to be biased towards the bright end of the magnitude distribution. How can we train in a semi-supervised fashion to take advantage of the unlabeled data? How can we  train using data from a different telescope with a different set of filters/cadences (i.e., transfer learning and domain adaptation)? How can we train models using synthetic or augmented data? How can we detect outliers in a stream of data? All of these are technically challenging problems which need to be developed, validated with the community, and then brought quickly into production.

Integration with the time--domain ecosystem and its community of users is another important challenge. First, brokers must be connected with other brokers, follow--up infrastructure, and data exploration tools. For this to happen, Application Programming Interfaces (APIs) must be developed, using Virtual Observatory (VO) or \emph{de facto} standards. Second, in order to produce relevant data products and tools, a frequent interaction with the community is needed to provide feedback and inject new ideas that can help improve the entire ecosystem. This includes interaction with small to large projects that interoperate with the community of survey telescopes, brokers, TOMs, and follow--up telescopes. A diversity of brokers must be encouraged, avoiding a winner--take--all solution, and fostering an environment where new, creative solutions rise faster into production.

\subsection{The ALeRCE Broker}

The ALeRCE broker is a Chilean-led project which aims to become a community broker for LSST and other large etendue survey telescopes. The project is run by an interdisciplinary team composed by astronomers, computer scientists and engineers, including faculty, postdoctoral fellows, and students. The broker's concept was first announced in 2017 as the natural continuation of the High cadence Transient Survey (HiTS), in which we used the Dark Energy Camera on the 4 m Blanco telescope to discover supernovae (SNe) in real--time by combining tools from high performance computing and ML \citep{2016ApJ...832..155F}. In 2018 a team of scientists was consolidated, the key requirements were defined, the first version of the front--end was developed, a memorandum of understanding was signed with the ZTF project, and the initial funding was secured. In early 2019, a dedicated team of engineers was hired to start building the tools needed to ingest the public ZTF alert stream in preparation for  LSST. 

ALeRCE started to systematically classify the ZTF stream using ML with astrophysically motivated taxonomies based on their light curves \citep{lateclassifier} since March 2019, and on their image stamps \citep{earlyclassifier} since July 2019. These classifiers are designed to balance the needs for a fast and simple classification with a subsequent, but more complex classification. ALeRCE has reported \nsncand SN candidates to the Transient Name Server\footnote{\url{https://wis-tns.weizmann.ac.il/}}, of which 361 have been spectroscopically confirmed. It has classified \nLCclf objects into a taxonomy that has expanded into 15 classes, including transient, periodic and stochastic variable sources, and with continuously improving precision and purity. All of ALeRCE's data products can be accessed freely via several dashboards, APIs, or a direct database connection.

 ALeRCE has adopted Agile work methodologies\footnote{\url{https://agilemanifesto.org/}}, which have been adapted to academic environments by several groups\footnote{\url{https://www.agilealliance.org/resources/experience-reports/reinventing-research-agile-in-the-academic-laboratory/}}. The main ideas behind these methodologies can be summarized as: 1) emphasizing individuals and interactions over processes and tools, 2) seeking improvements over sustaining practices, 3) collaboration over competition, and 4) adaptation to change over following a fixed plan. We use development \textit{sprints} of two weeks and short daily meetings where \textit{product owners} are the leading scientists of the different science cases, and where \textit{scrum masters} rotate among a few members of the team. It has been important to define precise and achievable objectives and associated deliverables in each sprint, coupling the team's skills and motivations around them. Adopting this methodology has important implications for the broker, which becomes a continuously evolving product with regular data and code releases. All the major components become dynamic: the classification taxonomy, as the available data sources grow and the product owners identify new scientific questions; the ML classification models, as new training sets and ideas are brought from development into production; and the tools and products, in order to adapt to the changing requirements of the community of users. This means that special attention needs to be given to version control of the broker pipeline, tools and data products. This is done via the use of GitHub repositories to track code changes, and the use of the Semantic Versioning\footnote{\url{https://semver.org/}} naming convention for our future pipeline and associated data releases, starting with version 1.0.0. 

The outline of this document is the following. In Section~\ref{sec:sciencegoals} we introduce the science goals of the ALeRCE broker, including a discussion of the broker taxonomy. In Section~\ref{sec:classifier} we describe the ML classifiers used by our broker. In Section~\ref{sec:pipeline} we present the pipeline structure and its associated infrastructure. In Section~\ref{sec:products} we discuss our main data products, services and tools. In Section~\ref{sec:results} we present some of the main results. Finally, in Section~\ref{sec:conclusions} we draw some conclusions and discuss future directions.

\section{Science goals} \label{sec:sciencegoals}

Our primary science goals are the study of three broad categories of objects: transients, variable stars and active galactic nuclei (AGN); we also provide Solar System object classifications as a secondary science goal.

\subsection{Transients}

Two important questions which can be answered via the study of transients are: 1) what is the nature of explosive phenomena, and 2) what can they teach us about the dynamics of the Universe. Rapid classification is key to answer these questions since it can facilitate dedicated follow-up observations, either rapid or slow, spectroscopic or photometric. Rapid follow--up is critical to understand short--lived transients and the progenitors of stellar explosions in general, since it probes the outermost, unprocessed layers of exploding stars and the possible interaction with the circumstellar medium \citep[e.g.,][]{2017NatPh..13..510Y, 2018NatAs...2..808F}. Early spectroscopy can be used to measure the composition and velocity structure of their ejecta. Late-time follow-up, either photometric or spectroscopic, probes the nature of the progenitor and explosion mechanism by constraining the composition and velocity structure of the innermost layers of the star \citep[e.g.,][]{2019NatAs...3..434F}. Having large samples of classified transient events cross--matched with multi--band/messenger or contextual information will help characterize the parameter space and provide clues of new, unrecognized populations of events. Furthermore, the ability to cross--match different streams in real-time, e.g., the LIGO and LSST streams, will offer possibilities which can lead to new, unexpected discoveries.  Finally, these larger and better calibrated samples, with well--understood systematics, can be used for cosmological distance and/or event rate estimations.

\subsection{Variable Stars}

 Some of the important questions which can be answered via the study of variable stars are: 1) what is the nature of these systems and the physical mechanisms of variability, and 2) what can they teach us about the structure and formation of our own galaxy, its satellites, and other galaxies in the Local Group \citep[e.g.,][and references therein]{2015pust.book.....C}. There are various reasons to obtain a uniform and rapid classification of variable stars. Rapid follow-up of stars entering/leaving the instability strip or changing their pulsation modes could provide new insights about the physics of stellar pulsation \citep[e.g.,][]{cg1999,bk2002,is2014}. Detection and follow-up of eclipses in pulsating stars can help provide direct stellar mass measurements \citep[e.g.,][]{ip2010,ip2012}. Rapid follow-up of gravitational microlensing events can allow the detection of planets with masses and separations resembling those in our Solar System \citep[e.g.,][]{br1996,agea2010}, while microlensing events with timescales of the order of years can provide clues about the nature of black holes (BHs) and dark matter \citep[e.g.,][]{2016PhRvD..94f3530G}. Moreover, microlensing may allow spectroscopic follow-up of sources that might otherwise have been too faint for spectroscopy \citep[e.g.,][]{bensby2020}. The detection of eruptive events and the spectroscopic follow-up immediately after the beginning of the eruption can provide new insights about the physics of young stellar objects \citep{2017MNRAS.465.3011C, 2018ApJ...861..145C}. Finally, larger and more distant samples of consistently classified variable stars \citep[e.g.,][]{2019A&A...623A.110G} will be key to understanding the tridimensional structure and formation history of our galaxy, along with that of its neighbors, ranging from the ultra--faint dwarfs to the Magellanic Clouds \citep[e.g.,][]{id2019,jd2020a,jd2020b,vivas2020}.

\subsection{Active Galactic Nuclei}

Some of the most exciting questions which can be answered from the study of AGN are: 1) what drives the growth of BHs \citep{2012NewAR..56...93A}; 2) what are the physical mechanisms behind AGN variability \citep{Sanchez-Saez18,2018MNRAS.480.4468R}; 3) are there intermediate-mass BHs \citep[IMBHs;][]{2017IJMPD..2630021M,2019arXiv191109678G}, with masses between stellar and super-massive BHs (SMBHs); 4) what is the structure and size of AGNs \citep{2016ASPC..505..107L}; and 5) what can tidal disruption events \citep{2014ApJ...793...38A} teach us about BH properties. Rapid classification could help identify and follow-up optical changing-look AGNs, a population which may unlock numerous clues to BH accretion physics \citep{LaMassa15,Graham19}. Selecting large samples of targets based on their multi--band variability for reverberation mapping studies can enable better physical constraints on the BH surrounding medium and distance \citep{2004ApJ...613..682P}. Fast cadence data can help assemble large samples of IMBHs candidates \citep{2020ApJ...889..113M}, which are known to vary on shorter timescales. The early detection of tidal disruption events can provide independent constraints on the BH properties that drive these phenomena \citep{Komossa15}. All of the above can be done while simultaneously cross-matching the LSST stream with future surveys that will provide critical additional information, such as {\it eROSITA} \citep{2012arXiv1209.3114M}, SKA precursors, IceCube \citep{ABBASI2009294}, etc. Finally, exploring larger samples of AGNs that are dimmer and redder can lead to the discovery of new populations of events and a better understanding of the AGN phenomena.

\section{ML classification} \label{sec:classifier}

\subsection{Classification Taxonomy} \label{sec:taxonomy}

An important component of an automatic classifier is the taxonomy used for classification, which defines the classes into which the alert stream will be classified. Choosing a good taxonomy is about achieving a balance between a reasonably accurate classifier, which depends on finding good training sets and the intrinsic separability of the classes, and meeting the demands of different communities of users. More complex taxonomies can be useful for a larger set of communities, but the addition of subclasses can lead to potentially less accurate classification models. The best compromise between the accuracy of the classifier and the complexity of the taxonomy is difficult to define, therefore in order to guide our choice of taxonomy we performed a survey of the taxonomies used in other studies that carried out ML classification of variable astronomical objects.

\subsubsection{Light Curve Classifier Taxonomy}

First, we consider those works that use only light curves in their analysis. We divide them into those that include both persistent variable and transient sources (Table~\ref{tab:MLboth}), those that include only persistent variable objects (Tables~\ref{tab:MLvarI} and \ref{tab:MLvarII}), and those that include only transient objects (Table~\ref{tab:MLtransients}). We examined four publications that include both transient and persistent variable objects in their taxonomy, 22 publications which include only persistent variable objects, and 8 publications which include only transient objects. There were 19 different sources of observational data, mostly for persistent variable sources (Table~\ref{tab:MLobssources}), and five sources of synthetic data (Table~\ref{tab:MLsyntheticsources}). 

A large diversity of taxonomies was found, with fewer classes in general being used in the last five years with respect to older works. This may be due to the appearance of more exploratory efforts in recent years, which look for variations from more traditional classification methods while using fewer classes for simplicity. We found more classes of persistent variable objects of stellar origin, probably because of the relative abundance of curated light curve training sets for these classes. The synthetic data sources were applied mostly for transient data, probably because of the relative difficulty in finding large numbers of observed transients. A brief description of the classes is included in the Appendix. The pulsating star variable classes included in the previous publications are shown in Tables~\ref{tab:variablesI} and \ref{tab:variablesII}, other stellar variable sources in Table~\ref{tab:variablesIII}, SMBH-related sources in Table~\ref{tab:AGNs}, and transients in Table~\ref{tab:transients}.

\begin{deluxetable*}{ccccc}[ht!]
\tablecaption{Light curve-based ML classifiers that include both transient and persistent variable objects. Note that \cite{lateclassifier} is an accompanying publication where we describe the ALeRCE Light Curve classifier in more detail.}
\label{tab:MLboth}
\tablehead{
\colhead{Reference}  & \colhead{Data source} & \colhead{Data type} & \colhead{\#classes} & \colhead{classes}\\
}
\startdata
\cite{lateclassifier} & ZTF & Observed & 15 & SNIa,~SNIbc,~SNII,~SLSN,\\
(See Section~\ref{sec:lateclassifier}) & & & & AGN,~QSO,~Blazar,~CV/Nova,~YSO,\\    
                 & & & & DSCT,~RRL,~Ceph,~LPV,~E, \\
                 & & & & Periodic--Other \\
\cite{2019AJ....158..257B} & PLAsTiCC & Simulated & 14 &     AGN,~RRL,~E,~Mira,~Mdwarf,~ML,\\
& & & & TDE,~kN,~SNIa,~SNIa-91bg,\\
& & & & SNIax,~SNIbc,~SNII,~SLSN-I\\
\cite{2018AJ....156..186M} & HiTS & Observed & 8 &  NV,~QSO,~CV,~SN,~DSCT,~E,~ROT,~RRL \\
\cite{2018ApJS..236....9N} & OGLE,OSC & Observed & 7 & SN,~BPer,~RRL,~LPV,~Ceph,~DSCT,~DPV\\
\cite{2016MNRAS.457.3119D} & CRTS & Observed & 6 & CV,~SN,~Blazar,~AGN,~Mdwarf,~RRL 
\enddata
\end{deluxetable*}

In general, there are certain families of objects which seem to be included consistently among most classifiers, but whose decomposition into subclasses varies greatly. Taking this into account we have decided to develop a hierarchical classifier which groups families of classes and which will gradually be refined as the amount and quality of the data grows \citep{lateclassifier}. The first level of the classifier considers transients, periodic, and stochastic variable phenomena. In the second level, the transient branch divides into (class names between parenthesis) the Type Ia SNe (SNIa), Type Ib and Ic SNe (SNIbc), Type II and IIn SNe (SNII), and Super Luminous SNe (SLSN) classes. The periodic branch divides into the eclipsing binary (E),  $\delta$ Scuti (DSCT), RR Lyrae (RRL), Cepheid (Ceph), long period variables (LPVs, including Miras, semi--regular and irregular variables), and other (Periodic-Other) classes. The Periodic-Other class corresponds to periodic objects which are not members of the E, DSCT, RRL, Ceph or LPV classes. The stochastic branch divides into host-dominated AGN, core-dominated AGN or quasi--stellar objects (QSO), blazars, cataclysmic variables and novae (CV/Nova), and young stellar objects (YSO).

\begin{figure*} [htb!]
\centering
\includegraphics[width=1\textwidth]{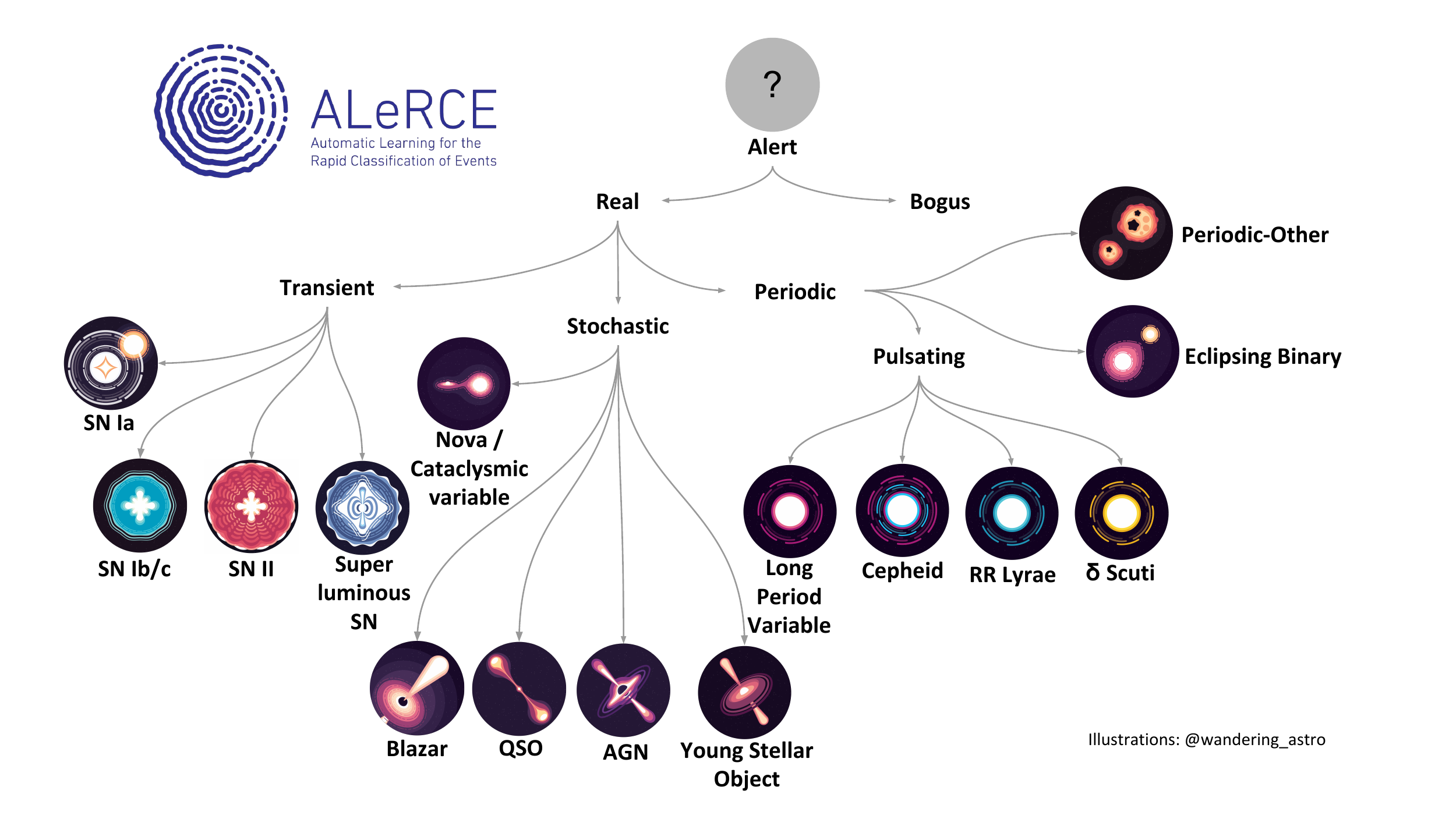}
\caption{\label{fig:taxonomy} The hierarchical taxonomy used by the ALeRCE broker for classifying light curves (v1.0.0). This classifier uses four models: one which separates transients, stochastic and periodic objects; another which separates transients into SNe Ia, SNe Ib/c, SNe II and Superluminous SNe; another which separates stochastic objects into blazars, QSOs, AGNs and YSOs; and another which classifies periodic stars into LPVs, Ceph, RRL, DSCT, Es or Periodic--Other.}
\end{figure*}

ALeRCE's current classification taxonomy is shown in Figure~\ref{fig:taxonomy}. This figure draws inspiration from the variability diagram of \cite{2008JPhCS.118a2010E}, most recently updated in \cite{leea2019}, but significantly simplified and with a more observationally based hierarchy, more resolution in the transient classes, and less resolution in the stellar variability classes. The reason for having more resolution in the transient classes is that in many cases the reaction time for the photometric or spectroscopic follow--up of these classes needs to be fast, e.g., to get spectroscopic confirmation or to characterize a short--lived phase of evolution, while for the persistent variability classes it is not as common to require fast follow--up. Thus, our main goal is to provide a first filter for the expert communities to explore further and classify into more complex taxonomies in more branches of the classification tree.

\subsubsection{Stamp Classifier Taxonomy} \label{sec:stampclassifiertaxonomy}

In addition to the classifiers which work solely on light curves, there are classifiers which use the pixel information contained on the variable object detection images. Alerts are generated from a difference image which results from aligning, scaling, convolving and subtracting the reference image from the science image. We have listed the ML classification studies which use the object ``image stamps" in Table~\ref{tab:MLstamps} for the classification of images into either real or bogus, but also as members of more astrophysically-motivated classes. The latter efforts are relevant for the taxonomy of our stamp--based classifier, a classification model which uses as input the first set of science, template and difference images associated with a new object in the alert stream\footnote{Note that the same object can have many associated alerts.}, and which is used as the first classification step in ALeRCE. Although the complexity of the taxonomy associated with this classifier is less refined, this early classification is critical to enable the triggering of fast photometric and spectroscopic follow--up and characterization of extragalactic transient sources. In the case of our stamp--based classifier \citep{earlyclassifier}, we have used the classes SN, AGN, variable star (VS), asteroid and bogus, trying to mimic how astronomers have historically looked for transients and variables. SNe tend to be near extended sources, AGNs are either relatively isolated point--like sources or at the center of extended sources depending on luminosity, variable stars are point--like sources which are frequently near other point--like sources and are present in both the science and reference images, asteroids are present only in the science image and not in the reference image, and bogus sources are not shaped like the point spread function of the image.

Finally, we found one publication that uses time series of image stamps \citep{2019PASP..131j8006C}, following an approach that combines time series and image stamps using a convolutional recurrent neural network classifier. They use seven classes: non--variable, galaxy, asteroid, SN, RRL, Ceph and E. This type of work could become more important in the future because it combines spatial and temporal information as well as simulated and real data.

\begin{deluxetable*}{cccc}[ht!]
\tablecaption{Single image stamp ML classifiers. Empirical data are used in all cases. Note that \cite{earlyclassifier} is an accompanying work where we describe the ALeRCE Stamp Classifier in more detail.}
\label{tab:MLstamps}
\tablehead{
\colhead{Reference}  & \colhead{Data source} & \colhead{\#classes} & \colhead{classes}\\
}
\startdata
\citealt{earlyclassifier} (Section~\ref{sec:earlyclassifier}) & ZTF & 5 & SN, AGN, VS, SN, asteroid, bogus \\
\cite{2019MNRAS.489.3582D} & ZTF & 2 & real,~bogus \\
\cite{2017MNRAS.472.1315W} & PanSTARRS1 & 3 & real,~asteroid,~bogus \\
\cite{2017ApJ...836...97C} & HiTS & 2 & real,~bogus\\
\cite{2017arXiv171111526K} & HSC-SSP & 2 & SNIa,~other \\
\cite{2015MNRAS.454.2026D} & SDSS & 2 & real,~bogus \\
\cite{2015AA...584A..44C} & RCS-2 & 2 & stars,~QSOs \\
\cite{2012PASP..124.1175B} & PTF & 2 & real,~bogus \\
\cite{2007ApJ...665.1246B} & PTF & 2 & real,~bogus \\
\enddata
\end{deluxetable*}

\subsection{Training Sets}

In order to compile training sets, we use only sources observed by ZTF whose labels have been cross--matched from different catalogs available in the literature, or compiled by our collaboration. For each catalog we define a function which maps the catalog's taxonomy into our own taxonomy, allowing us to aggregate labels from different catalogs into a unified taxonomy. Then, we assign a priority order that defines which labels to use in case of disagreement between catalogs. These priorities are based on discussions with community experts, a critical analysis of the methods that were used to classify objects (e.g., manual vs. automatic), and an analysis of which catalogs tend to disagree more with other catalogs, from a visual exploration of catalog label matrices (similar to confusion matrices, but with rows and columns as the classes in each catalog, potentially with different taxonomies).

The catalogs we use to extract labels from are, in order of priority: 

\begin{enumerate}
    \item Cataclysmic variables catalog: compiled by \cite{Abril20}, including \citealt{Ritter03}.
    \item ROMABZCAT: Multi--frequency catalog of blazars from \cite{2015Ap&SS.357...75M}.
    \item Catalog of Type I AGNs from \cite{2015ApJS..219....1O}.
    \item The Million Quasars (MILLIQUAS) Catalogue from \cite{2019arXiv191205614F}.
    \item Spectroscopically classified SNe in the Transient Name Server, TNS.\footnote{\url{https://wis-tns.weizmann.ac.il/}}
    \item Objects classified as YSOs in Simbad \citep{2000A&AS..143....9W}.
    \item Catalina Real Time Transient Survey (CRTS) catalog of northern periodic sources \citep{2014ApJS..213....9D}.
    \item CRTS catalog of southern periodic sources \citep{2017MNRAS.469.3688D}.
    \item The LINEAR catalog of periodic variables \citep{2013AJ....146..101P}.
    \item {\em Gaia} Data Release 2 (DR2) catalog of variable stars \citep{2018A&A...618A..58M}.
    \item The ASAS-SN catalog of variable stars \citep{2019MNRAS.486.1907J}.
\end{enumerate}

\subsection{The Light Curve Classifier} \label{sec:lateclassifier}

This classifier computes classification probabilities for objects with $\geq6$ detections in $g$ or $\geq6$ detections in $r$. We represent individual light curves as a vector of features compiled from the literature and new features developed by the ALeRCE collaboration as described in \citet{lateclassifier}. One of the most relevant new features comes from an irregularly sampled autoregressive model (IAR) introduced in \cite{2018MNRAS.481.4311E}, which is able to estimate autocorrelation in irregularly sampled time series in a statistically robust way. The classification is done in a hierarchical fashion using a balanced random forest classifier\footnote{Using the \href{https://imbalanced-learn.readthedocs.io/en/stable/api.html}{imblearn} library}, which in our tests achieved better accuracies than recurrent neural networks \citep[e.g.,][]{2019PASP..131k8002M}. As described before, a given object will be first classified as either periodic, stochastic or transient and subsequently refined into 15 different classes as described in Section~\ref{sec:taxonomy}. The latest confusion matrix associated with this classifier can be seen in Figure~\ref{fig:lateconfusion}, described in  \cite{lateclassifier}.

\subsection{The Stamp Classifier} \label{sec:earlyclassifier}

Inspection of ZTF image stamps suggests that it should be possible to classify alerts based on the first detection set of stamps (see Section~\ref{sec:stampclassifiertaxonomy}). Therefore, we designed and trained a stamp classifier based on a convolutional neural network with the main motivation of finding SN candidates using as input the information contained in the first alert, including the science, reference and difference stamp set, as well as other meta data, such as spatial location and data quality metrics.

\begin{figure}[htbp]
\begin{center}
\includegraphics[width=0.45\textwidth]{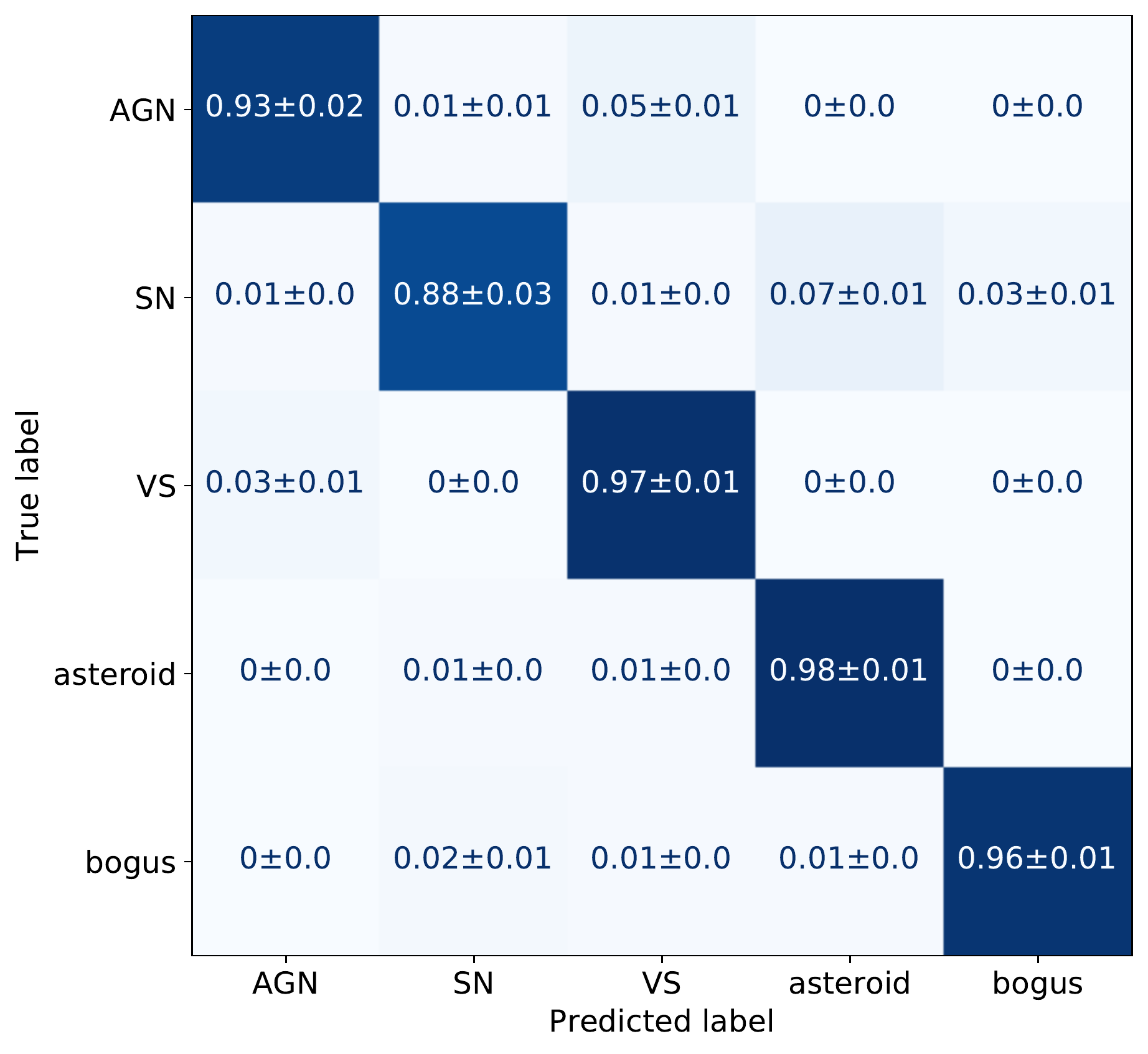}
\caption{Confusion matrix obtained with the stamp classifier model in \cite{earlyclassifier}. \label{fig:earlyconfusion}}
\end{center}
\end{figure}

The stamp classifier \citep{earlyclassifier} is able to discriminate among five classes: SNe, AGN, variable stars, asteroids, and bogus alerts, achieving 90\% accuracy on a balanced test set, and a recall of 81\% among spectroscopically confirmed SNe from TNS. To improve the model interpretability, we added a regularization term that maximizes the entropy of the predicted probability for each class, enhancing the different certainties for each prediction. This model is currently running on ZTF alerts and its results are publicly available in the ALeRCE SN Hunter at \url{https://snhunter.alerce.online} (see Section~\ref{sec:web}). The confusion matrix associated with this classifier can be seen in Figure~\ref{fig:earlyconfusion}, reproduced from \cite{earlyclassifier}.

\subsection{Metrics and Selection of Classification Model}

In order to evaluate the classifiers that will go from initial model training into production, we use a combination of metrics and tests that take into account the labeled and unlabeled data. We have found this to be relevant when using a labeled training set known to be non--representative of the unlabeled data. First, we compute the test set classification balanced (averaged per class) accuracy (ratio between correct and total labels), and F1--score (the harmonic mean between precision and recall) to take into account the accuracy, precision and recall of the classifier while considering the class imbalance, which is very important when using observational data as training sets. Second, we look at the confusion matrix to search for signs of over--representation of certain classes which may not be evident in the balanced accuracy. Third, for the light curve classifier we look for classification biases with certain relevant variables; e.g., looking for a relatively constant recall vs. apparent magnitude relation for individual classes when no significant bias exists. Fourth, we compare the expected and inferred spatial and class distributions of the unlabeled data to discard models using astrophysical knowledge. For example, if the classification model were correct one would expect the spatial distribution of the different classes to follow known patterns, such as that most Galactic classes should be concentrated around the Galactic plane, extragalactic classes should be homogeneously distributed outside the Galactic plane due to extinction and source confusion, and asteroids should be distributed around the ecliptic. Additionally, we would expect the distribution of class labels in the unlabeled set to follow known population ratios, for example we expect SNe Ia to be more abundant than SNe Ibc. Therefore, the final choice of a classification model is made considering all these metrics and tests before the model is brought into production, i.e., applying the model using the available infrastructure with our latest pipeline for nightly operations.

\begin{figure*}[htbp]
\begin{center}
\includegraphics[width=0.8\textwidth]{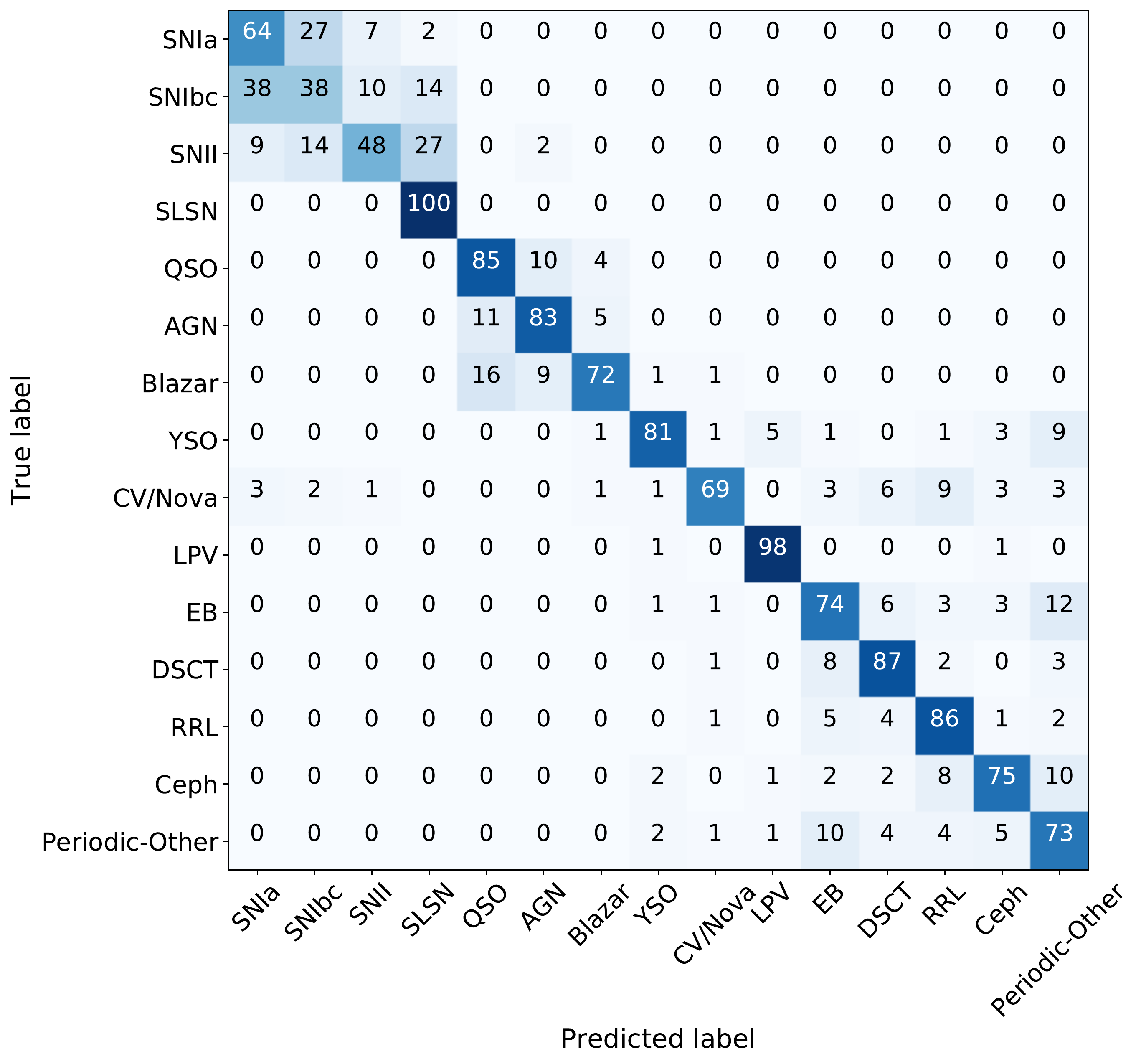}
\caption{Confusion matrix obtained with the balanced hierarchical random forest light curve classifier model in \cite{lateclassifier}. \label{fig:lateconfusion}}
\end{center}
\end{figure*}

\subsection{Stamp and Light Curve Classifier Comparison}

\begin{figure*}[htbp]
\begin{center}
\includegraphics[width=1\textwidth]{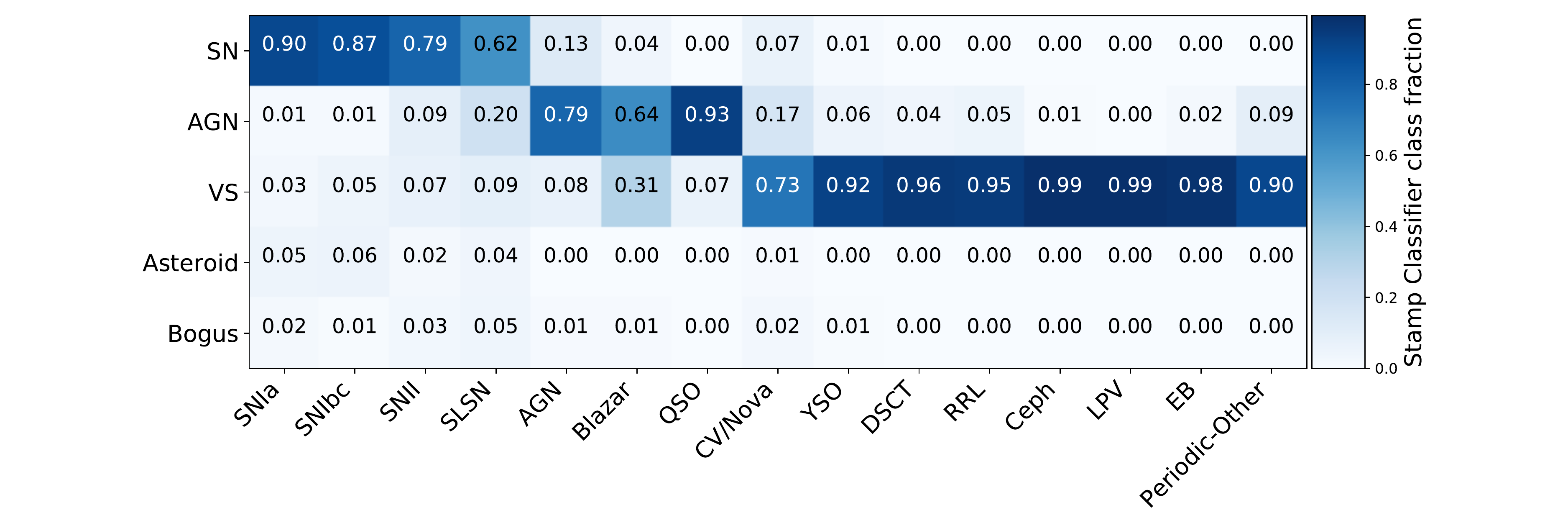}
\caption{Fraction of objects predicted to belong to a given Stamp Classifier class (rows), normalized among the objects predicted to belong to a given Light Curve Classifier class (columns). We considered a sample of 186,794 unlabeled objects which were classified with the Stamp Classifier \cite{earlyclassifier} and the Light Curve Classifier \cite{lateclassifier}. \label{fig:stampvsLC}}
\end{center}
\end{figure*}

As a consistency check between the two aformentioned classifiers, we compare the distribution of classes of the Stamp Classifier among those objects classified by the Light Curve classifier. In Figure~\ref{fig:stampvsLC} we show a matrix of Stamp Classifier classes and Light Curve Classifier classes, normalized along the Light Curve Classifier classes. We can see that there is overall agreement between the two classifiers, which highlights the complementarity between our two classifiers, and emphasizes the value of using the image stamps for early classifications as shown in \cite{2019PASP..131j8006C}.

\subsection{Outlier/Novelty Detection} \label{sec:outlier}

Outlier/novelty detection refers to the automatic identification of abnormal or unexpected phenomena embedded in data \citep{Faria2016}. We are developing outlier detection methods experimentally to focus on two problems: the discrimination of outlier clusters of time series or image stamps, i.e., cohesive and representative sets of examples associated with interesting phenomena that are not characterized in the current training database; and the detection of unexpected events occurring within a particular time series. To solve the first problem we are developing online one--class/semi--supervised outlier detection methods \citep{oneclassoutlier, Chapelle2009, 2020arXiv200507779R} to find similarities between objects and automatically detect outlier phenomena. We are addressing this problem from three different perspectives: using autoencoders, generative adversarial networks, and one--class neural networks. To find unexpected events within time series, we are using robust online nonlinear filters \citep{Liu2011,Huentelemu2016}. Traditional methods such as Kalman filters and kernel filters are being extended to incorporate measurement uncertainties, the heteroscedasticity of the noise, and the use of state space formulations where states are unevenly separated in time.

For both problems, Active Learning techniques \citep{Zhu2003} are being explored to select sets of the most uncertain objects and/or events to be shown to human experts. We are aiming to use information theoretic feature selection \citep{Estevez2009} and feature extraction methods to reduce dimensionality and generate visualizations that can be presented to the experts. 

\section{ALeRCE pipeline and infrastructure} \label{sec:pipeline}

\begin{figure} [htb] 
\centering
\includegraphics[width=0.51\textwidth]{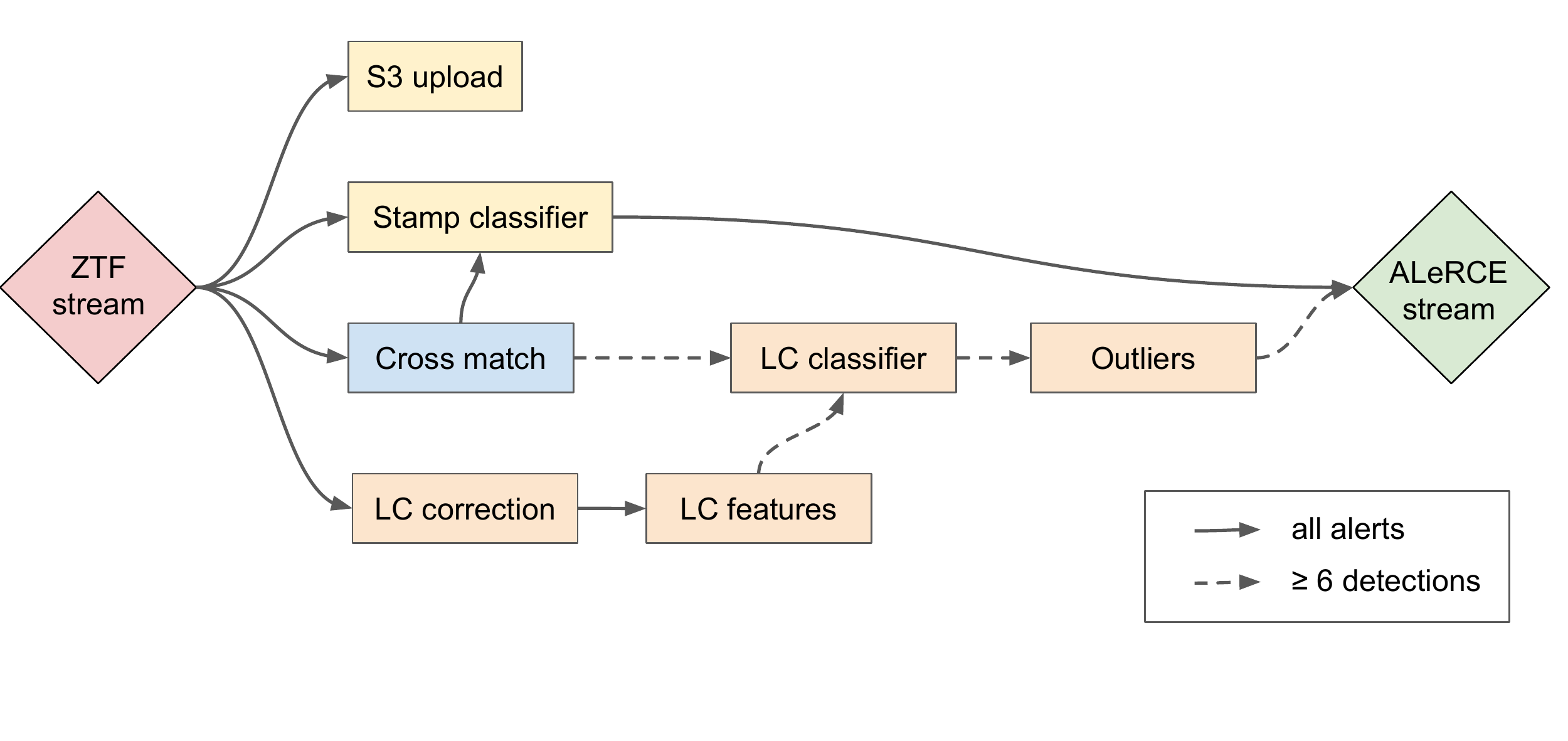}
\caption{ALeRCE pipeline structure from ZTF alert ingestion to the ALeRCE streaming of the processed alert. Alerts ingested from the public ZTF stream are first sent to four parallel Kafka topics: an Avro backup service in AWS S3, the stamp classifier for early SN detections, a cross--match step to gather information from public catalogs, and a light curve (LC) correction step. The LC correction step is followed by a LC features computation step, and a LC classifier and outlier detection steps, which are only applied to objects with 6 or more detections. Note that the ML classification steps can also be fed with information from the crossmatch step. The tables of our database are modified inside the pipeline steps for subsequent access via APIs. \label{fig:pipeline}}
\end{figure}

ALeRCE is currently processing the alert stream provided by the ZTF survey, but we expect to ingest other alert streams in the future, such as those provided by ATLAS, HATPi\footnote{\url{https://hatpi.org/science/}} and LSST (see Figure~\ref{fig:etendue}). The ZTF pipeline and alert distribution system are described in \cite{2019PASP..131a8003M} and \cite{2019PASP..131a8001P}. Alert packets contain image difference stamps and other metadata, whose detailed description can be found in \url{https://zwickytransientfacility.github.io/ztf-avro-alert/schema.html}. The ALeRCE system ingests these alerts and processes them through a pipeline which is divided into a combination of sequential and parallel steps, shown schematically in Figure~\ref{fig:pipeline} and described below.

\subsection{Ingestion and Kafka Topics}

ZTF alerts are sent as Avro packets\footnote{\url{https://avro.apache.org}} which contain associated image stamps, metadata and information related to previous detections as described in \url{https://zwickytransientfacility.github.io/ztf-avro-alert/schema.html}. We use Apache Kafka\footnote{\url{https://kafka.apache.org}} to receive the ZTF alert stream and to communicate information between the different steps of our pipeline as independent Kafka topics. We use an \href{https://zookeeper.apache.org/}{Apache Zookeeper} cluster with a replication factor of three, following recommended practices, and three independent machines of Kafka consumers, which are responsible for reading data from the alert queue. We have set up a Kafka cluster in Amazon Web Services (AWS) to manage different topics associated with different steps in the pipeline. Assigning different topics for each step in the pipeline has the advantage of allowing for alerts to be grouped in different batch sizes optimized for performance. For example, querying the database for several objects simultaneously can be faster than doing it sequentially for a list of objects depending on the type of query, or in the case of cross--matching, it may be more efficient to group alerts by their spatial location if the external catalog is stored hierarchically, e.g., a tessellation of the sky. Another advantage is that we can configure each topic independently for performance, e.g., using different numbers of Kafka partitions per topic.

We have tested different configurations of Kafka producers to mimic an LSST--like stream of data, and we have found that a cluster of three Kafka consumers with 12 partitions each is capable of ingesting all the different topics at a rate of 119.7 MB/s, which is about three times faster than the average alert production rate expected for LSST.

\subsection{Database and Avro Repository} \label{sec:database}

As alerts arrive, we store the original Avro files in AWS Simple Storage Service (S3) buckets for future analysis and extract a selection (in order to limit the size of the database) of the fields contained in these packets to be added directly to a database using a PostgreSQL database engine. As the data are processed and object alerts aggregated, we add different statistics to different tables. The main tables in our database are: 
\begin{itemize}
    \item \verb+objects+ table, which contains basic filter and time--aggregated statistics such as location, number of observations, and the times of first and last detection.
    \item \verb+magstats+ table, which contains time--aggregated statistics separated by filter, such as the average magnitude, or the initial magnitude change rate.
    \item \verb+detections+ table, which contains the object light curves including their difference and corrected magnitudes and associated errors separated by filter (see Section~\ref{sec:LCcorrection}).
    \item \verb+non_detections+ table, which contains the limiting magnitudes of previous non--detections separated by filter.
    \item \verb+features+ table, which contains the object light curve statistics and other features used for ML classification and which are stored as json files in our database.
    \item \verb+xmatch+  table, which contains the object cross--matches and associated cross--match catalogs.
    \item \verb+classification+ tables, which contain the object classification probabilities, including those from the stamp and light curve classifiers, and from different versions of these classifiers.
    \item \verb+taxonomy+ table, that contains details about the different taxonomies used in our stamp and light curve classifiers, which can evolve with time. 
\end{itemize}
     
A webpage containing an updated description of the different tables can be found in \url{https://alerce.science}. As the volume of alerts grows for different projects, we expect to migrate some of the previous tables to NoSQL database engines such as Cassandra or MongoDB. After ingestion, the alerts undergo the processing steps described next.

\subsection{Stamp Classification}

When an alert from a previously unreported  object arrives, its first available image stamps are used to classify it as either SN, AGN, variable star, asteroid or bogus, as explained in Section~\ref{sec:earlyclassifier}. Note that if the first detection from an object did not pass the ZTF real/bogus test, but a subsequent detection did, the first available image stamp will not be from the former.  This stamp classification is done within one second of the alert being received and is automatically available in our database and in the SN Hunter tool (see Section~\ref{sec:web}), if the candidate is consistent with being a SN. The details of the stamp classifier are described in a parallel publication \citep{earlyclassifier}.

\subsection{Light curve Correction} \label{sec:LCcorrection}

As explained before, ZTF alerts are produced when a science image contains a significant change with respect to a reference image, after aligning, scaling, convolving and subtracting the reference image from the science image. Flux differences with respect to the reference image are reported as difference magnitudes and an associated flag (\verb+isdiffpos+) is included to indicate whether the difference is positive or negative.  In the case of ZTF, a reference image is defined by a unique reference field identifier (\verb+rfid+). If the source was present in the reference image it is possible to recover its actual apparent magnitude from the difference and reference magnitudes. We do this correction when the nearest catalogued object is closer than 1.4" (\verb+distnr+$<1.4$), providing a flag to indicate whether we think the object is extended based on PanSTARRS and ZTF shape parameters. The actual apparent magnitude and associated errors in the case of an point--like source which was present in the reference are the following:
\begin{eqnarray}
    m_{\rm corr} &= -2.5 \log_{10} \bigl( 10^{-0.4 ~m_{\rm ref}} + {\rm sgn} ~ 10^{-0.4~ m_{\rm diff}} \bigr) \\
    \delta m_{\rm corr} &= \frac{\bigl(10^{-0.8 m_{\rm diff}} \delta m_{\rm diff}^2  \bigl[ - 10^{-0.8 m_{\rm ref}} \delta m_{\rm ref}^2 \bigr] \bigr)^{0.5}}
        {10^{-0.4~ m_{\rm ref}} + {\rm sgn}~ 10^{-0.4 ~m_{\rm diff}}} \label{eq:corr}
\end{eqnarray}
where $m_{\rm ref}$ is the magnitude of the object in the reference image, $m_{\rm diff}$ is the magnitude associated with the absolute flux difference between the science and reference images, ${\rm sgn}$ is the sign of the difference (\verb+isdiffpos+), $\delta m_{\rm ref}$ is the error associated with the reference magnitude, and $\delta m_{\rm diff}$ is the error associated with the difference magnitude. Note that we provide both the original and corrected photometry. For the corrected photometry, we include errors values with and without the term inside square brackets in Equation~\ref{eq:corr}, which originates from the correlation between the reference and difference fluxes (see derivation in Appendix~\ref{sec:LCcorrection_appendix}).

It is important to note that if the difference flux is equal to the reference flux and the sign of the difference is negative, both the corrected magnitude and associated errors will diverge, which is a limitation of using a logarithmic scale for difference fluxes. This should normally not occur, since an alert is triggered only when there is a significant difference with respect to the reference. However, if the reference image contains a transient source, the difference flux can eventually become exactly minus the reference flux, and the corrected flux zero, which will lead to divergences depending on the noise. We treat these cases by assigning values of 100 to the corrected magnitudes and their associated errors.

We discuss in detail the derivation of these formulae, how to include the effect of a change in reference image, and how we treat extended sources in the reference image in the Appendix~\ref{sec:LCcorrection_appendix}. 

\subsection{Xmatch}

A cross-match step runs in parallel with the stamp classifier and light curve correction, querying external catalogs in order to extract additional information about the objects of interest. The ZTF alert packets already contain the nearest Solar System, PanSTARRS and {\em Gaia} catalogued sources. In addition to this information, we query WISE and SDSS in order to obtain infrared  and spectroscopic information if available, which can be critical to better constrain some of the classes included in our taxonomy. Additional catalogs will be included as they prove relevant. These queries are done using the CDS cross--match API\footnote{\url{http://cdsxmatch.u-strasbg.fr/xmatch/doc/}}.

\subsection{Feature Computation}

With the corrected light curves we can compute light curve characteristics or features based on both the detections and non--detections of a given object, but also on available crossmatches. Advanced light curve features are only triggered for objects with $\geq6$ detections in $g$ or $\geq6$ detections in $r$. The features computed are a significantly extended version of the FATS library \citep{2017ascl.soft11017N}, called Turbo FATS, which is optimized for computation speed and adds several new features. A description of these features, which are contained in the \verb+features+ table of our database, can be found in \cite{lateclassifier}.

\subsection{Light Curve Classification}

Objects having computed features are then processed by the light curve classifier described in Section~\ref{sec:lateclassifier}. The results of this classifier are obtained within a few seconds from ingestion for 95\% of the objects. For a larger stream this could be maintained by scaling the infrastructure given the \textit{embarrassingly parallel} nature (i.e., no need of communication between parallel tasks) of the light curve correction, feature computation and light curve classification tasks between different alerts. The current model used for the light curve classifier is a hierarchical balanced random forest, as described in \cite{lateclassifier}. 

After the light curve classification step we perform an outlier detection step, which as of Jun 2020 is being actively developed experimentally (see Section~\ref{sec:outlier}).

\subsection{Database Integrity Tests}

After the nightly ingestion and processing of the alerts, we perform a series of database integrity tests during the day. This consists in reanalyzing the Kafka topic associated with the last night of observations to check that no alerts were lost during the processing due to unexpected errors. If any alerts were missed during the night, we add them to a specially created Kafka topic which is then processed by our pipeline until no missing alerts exist.

\section{Data Products and Services} \label{sec:products}

The ALeRCE broker provides several data products and services which are constantly growing as we identify new requirements from our community of users. New requirements are defined by \textit{user stories}, informal descriptions of desired features from the perspective of an end user, which are translated into different data products and services by astronomers in our team following an Agile methodology. In this section we list the most important data products and services provided by ALeRCE as of Jun 2020, which are summarized in Table~\ref{tab:dataproductsservices}.

\begin{table*} 
\centering
\caption{Summary of ALeRCE data products \& services as of Jun 2020.} \label{tab:dataproductsservices}
\begin{tabular}{ c c c}
Type & Name & Address \\
\hline
Database & ALeRCE DB PostgreSQL repository  &
\url{db.alerce.online} \\
\hline
GitHub repositories & ALeRCE open source repositories & \url{http://github.com/alercebroker} \\
\hline
Jupyter notebooks & Science use cases notebooks & \url{http://github.com/alercebroker/usecases} \\
Jupyter notebooks & TNS upload notebooks & \url{http://github.com/alercebroker/TNS_upload} \\
\hline
Output stream & ALeRCE output Kafka stream & \emph{Please contact us.} \\
\hline
Website & ALeRCE main webpage & \url{http://alerce.science/} \\
\hline
Dashboard & ALeRCE Grafana pipeline dashboard\footnote{Request access} & \url{http://grafana.alerce.online/} \\
\hline
Documentation & ALeRCE API documentation & \url{http://alerceapi.readthedocs.io/en/latest/} \\
Documentation & ALeRCE client documentation & \url{http://alerce.readthedocs.io/en/latest/} \\
Documentation & ALeRCE tutorial videos & \url{https://bit.ly/2NHDagc} \\
\hline
Web interface & ALeRCE explorer & \url{http://alerce.online} \\
Web interface & SN Hunter & \url{http://snhunter.alerce.online} \\
Web interface & Crossmatch interface & \url{http://xmatch.alerce.online} \\
Web interface & ALeRCE reporter & \url{http://reporter.alerce.online/} \\
Web interface & TOM Toolkit plugin & \url{http://tom.alerce.online/} \\
\hline
API & ZTF DB access & \url{http://ztf.alerce.online}\\
API & Avro/stamp service & \url{http://avro.alerce.online}\\
API & ZTF crossmatch service & \url{http://xmatch-api.alerce.online}\\
API & catsHTM crossmatch service & \url{http://catshtm.alerce.online}\\
API & TNS crossmatch service & \url{http://tns.alerce.online}\\
API & Finding chart generator & \url{http://findingchart.alerce.online}\\
\hline
\end{tabular}
\end{table*}

\subsection{Data Products}

The ALeRCE data products can be divided into several categories: the tables of a database, a repository of Avro files, a repository of jupyter notebooks, an output stream of annotated and classified alerts, a GitHub repository with our open source code, a Grafana dashboard to monitor the status of the pipeline, our main webpage, documentation webpages, and tutorial videos for new users. We provide a brief description of each of them in what follows. 

\subsubsection{Database}

The tables in our database integrate the information about individual objects. A description of the database can be found in Section~\ref{sec:database}. The tables from our database are open for direct exploration in read--only mode as shown in some of our use case jupyter notebooks (\url{https://github.com/alercebroker/usecases}), although we recommend accessing them using our different APIs for simple queries (see Section~\ref{sec:API}). A detailed description of the tables and schema used in our database can be found in \url{http://shorturl.at/cJS34}.

\subsubsection{Avro Repository}

Apart from the previous tables, a copy of the original Avro files contained in the ZTF stream are stored in AWS S3. These Avro files can be accessed using our Avro/stamp API.

\subsubsection{GitHub Repositories}

All of our open source code can be found in the GitHub repository \url{https://github.com/alercebroker}. In the course of developing this project and as of Jun 2020 we have created 113 repositories, 27 of which have been made public for our community of users. These repositories can be forked or modified for external use. The pipeline steps are contained in these repositories and new version numbers are defined when dockerized versions of the steps are created.

\subsubsection{Use Case Jupyter Notebooks}

We have compiled a list of example jupyter notebooks which show how to use our API or directly access our database, focused around different science cases, such as SN, variable stars, AGN, or even asteroid studies. They can be found at \url{https://github.com/alercebroker/usecases}.

Apart from these notebooks, we have created a special notebook and associated GitHub repository for the inspection and submission of SN candidates to TNS (\url{https://github.com/alercebroker/TNS_upload}). In this notebook users can interact with Hierarchical Progressive Surveys \cite[HiPS,][]{2015A&A...578A.114F} PanSTARRS images to easily select the candidate host galaxies using \href{https://github.com/cds-astro/ipyaladin}{ipyaladin}, \href{https://ned.ipac.caltech.edu/}{NED}, \href{http://simbad.u-strasbg.fr/simbad/}{Simbad}, and \href{https://www.sdss.org/dr15/}{SDSS DR15}. This repository includes a tutorial explaining all the steps required to upload candidates to TNS, including tutorial videos to guide users in the process.

\subsubsection{Output Stream}

A real--time output stream is provided to report database changes as new alerts arrive and are processed by our pipeline, including an update on the classification probabilities and basic statistics. Users can connect to this stream using Apache Kafka upon request.

\subsubsection{Grafana Dashboard}

A Grafana dashboard is available to monitor the ALeRCE pipeline and associated database and infrastructure (\url{http://grafana.alerce.online}). This dashboard shows the status of the Apache Kafka servers and relevant metrics about the number of alerts being processed, the PostgreSQL database and associated servers, and the front--end servers. Access to this dashboard can be given upon request.

\subsubsection{Main Website, Documentation and Tutorial Videos}

ALeRCE's main website, which summarizes all our data products and services, can be accessed at \url{http://alerce.science}. Documentation for our API services and client (see Section~\ref{sec:web}), and a series of tutorial videos for our community of users can be found at \url{https://bit.ly/2NHDagc}.

\subsection{Services}

Apart from the previous data products, several services are provided to facilitate the exploration of the ZTF stream and associated objects. They are divided into web interfaces, which are web pages that allow the simple exploration of the alert stream; and APIs, which power the previous web interfaces and allow for the flexible integration of ALeRCE into the time domain ecosystem.

\subsubsection{Web Interfaces}  \label{sec:web}

\begin{figure} [htb] 
\centering
\includegraphics[width=0.45\textwidth]{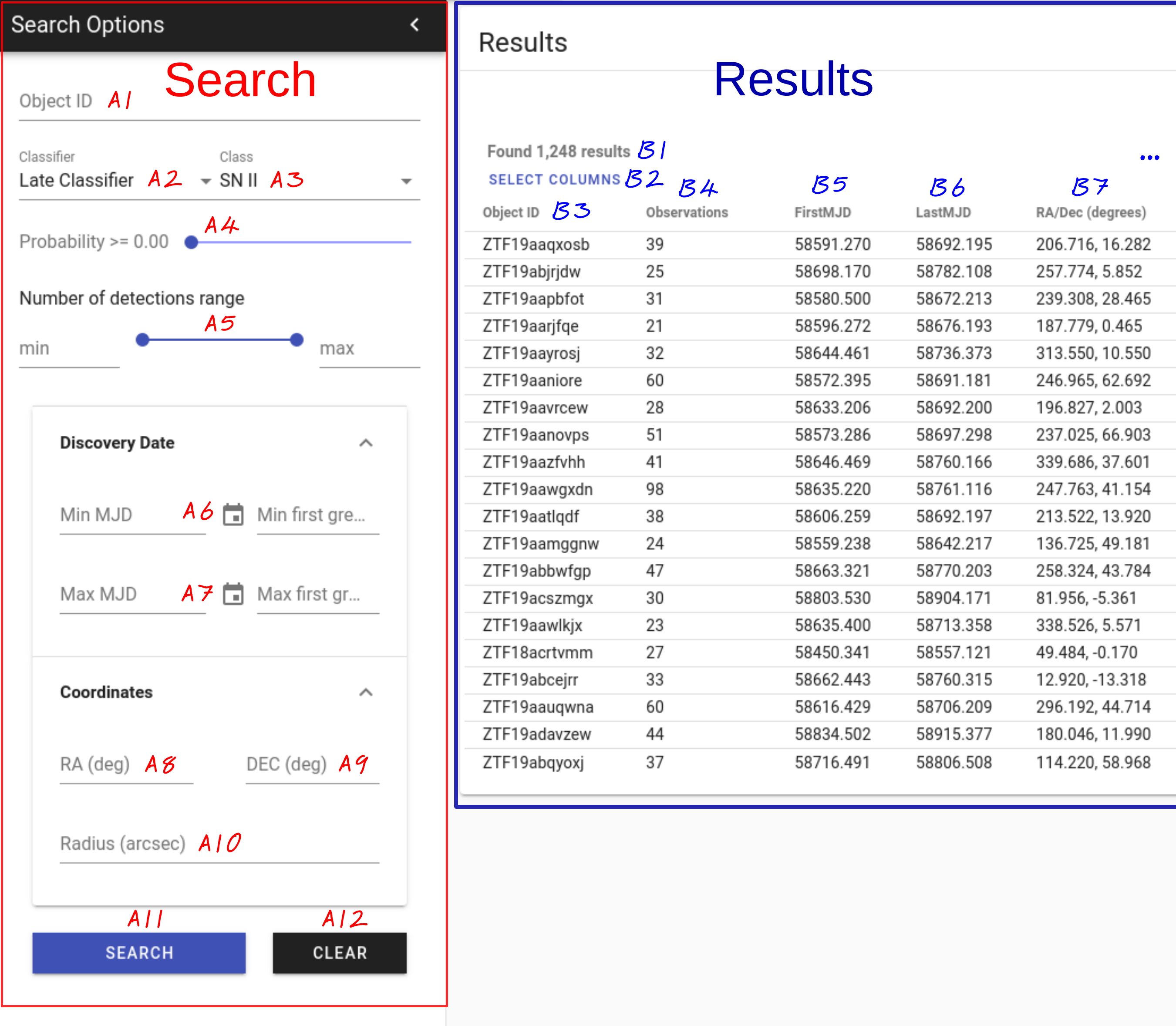}
\caption{The ALeRCE explorer web interface (\url{http://alerce.online}) initial \textbf{Search and Results} view. The \textbf{Search} panel allows users to directly filter by object identifier (A1); by inferred type using either the stamp or light curve classification  models (A2), a given class (A3), and a minimum classification probability (A4); by the minimum and maximum number of detections (A5); by minimum (A6) and/or maximum (A7) discovery date in modified Julian dates or calendar dates; or by location in the sky using a cone search defined by a right ascension (A8), declination (A9), and search radius (A10). The Search button (A11) submits queries and the Clear button (A12) clears the search options. The \textbf{Results} panel shows the results of the previous query. First, it shows the total number of results (B1), which are displayed in a paginated format. Users can select which columns to display (B2). The columns shown in this figure are the object identifier (B3), the number of detections (B4), the time of first (B5) and last (B6) detection, and the coordinates (B7). Other columns displayed by default (not shown in this image) are whether the object has cross--matches, and the stamp and light curve classifier classes and probabilities.  Clicking on an object links to the Object view (Figure ~\ref{fig:explorer_2}). \label{fig:explorer_1}}
\end{figure}

\begin{figure*} [htb]
\centering
\includegraphics[width=1.0\textwidth]{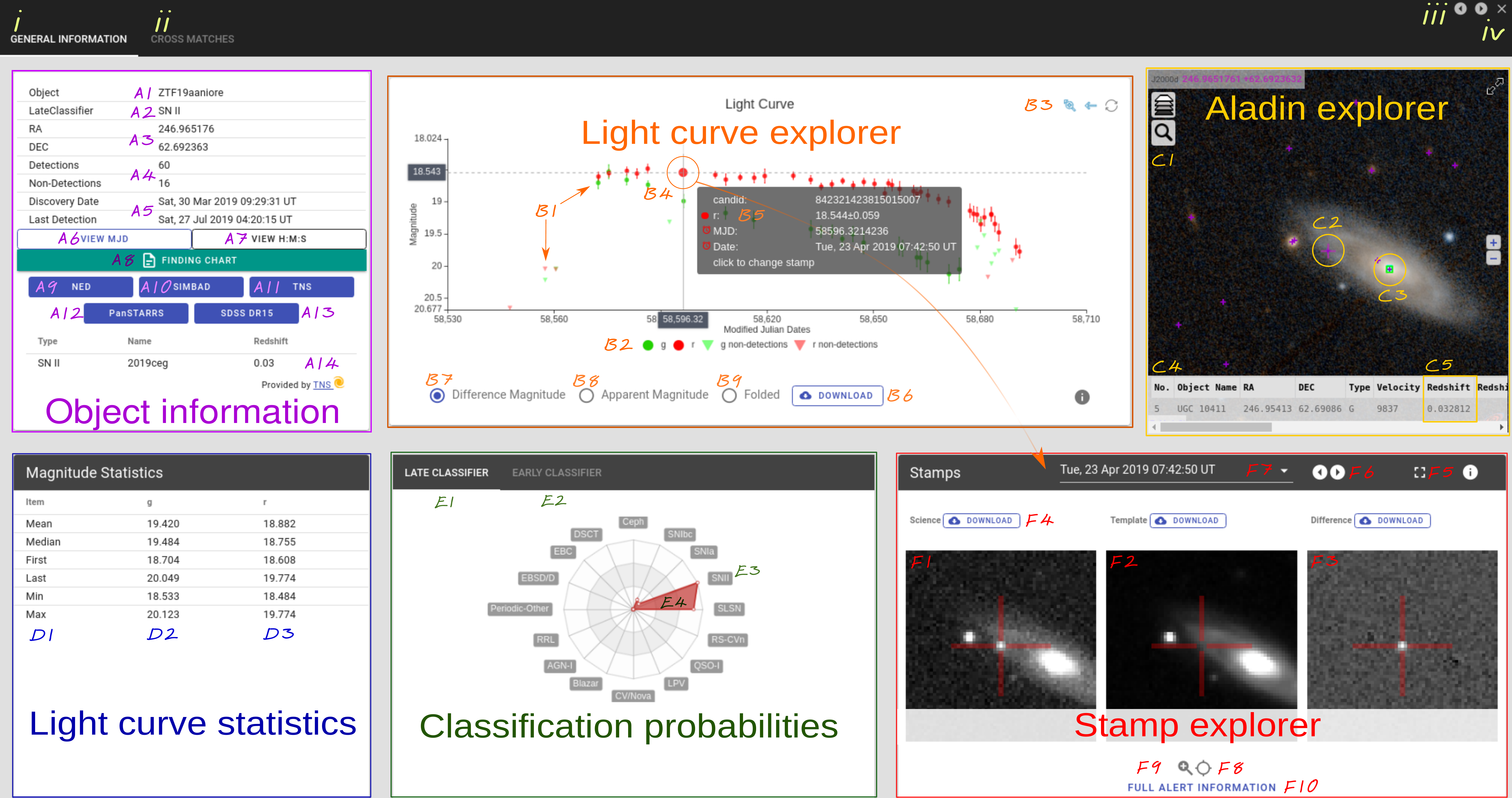}
\caption{The ALeRCE explorer web interface (\url{http://alerce.online}) object view. At the top left, users can switch between the General information (\textit{i}, this figure) or the Cross--matches (\textit{ii}) object views, and at the top right, between different objects (\textit{iii}, or the arrow keys) if directed from the results table of a previous query, or to go back to the Search and Results view (\textit{iv}, or the escape key). The \textbf{General information view} contains six different panels which we demarcate with colored text. At the top left, the \textbf{Object information} panel shows the object's unique identifier (A1); most likely class (A2); coordinates (A3) in different formats (A7); number of detections and non--detections (A4); and the first and last detection times (A5) in calendar or modified Julian dates (A6). It also contains links to the finding chart generator tool (A8); the NASA Extragalactic Database (A9, NED); Simbad (A10); TNS (A11); PanSTARRS (A12); and the SDSS DR15 navigation tool (A13). The latest type, name and redshift associated with the object in TNS are also shown (A14). At the top middle, the \textbf{Light curve explorer} panel displays the latest light curve of the object, including both detections and non--detection upper limits in both bands (B1), which can be turned on/off individually (B2). The light curve can be zoomed in and out (B3) and users can hover over individual points (B4) to see the exact date, magnitude and alert identifier (B5), or click on to display its associated stamps and full alert information in the Stamp explorer panel. The light curve and associated data can be downloaded (B6) and users can select whether to show: the difference magnitude (B7); the apparent magnitude (B8), i.e., corrected by the flux in the reference image; or the period--folded apparent magnitude (B9), assuming that the light curve is periodic and using a periodogram to compute the most likely period. At the top right, the \textbf{Aladin explorer} panel shows an interactive Aladin window (C1) with a PanSTARRS image at the location of the candidate (C2), in this case a confirmed SN near its likely host galaxy (C3). An overlaid catalog of objects can be clicked on to view more information (C4), such as the host redshift (C5). At the bottom  left, the \textbf{Light curve statistics} panel shows different statistics (D1) computed over the $g$ (D2) and $r$ (D3) bands of the apparent magnitude light curve. At the middle bottom, the \textbf{Classification probabilities} panel shows the classification probabilities according to our light curve (E1) or stamp (E2) classifiers, when available. A radar plot of the class probabilities for the taxonomy used in the classification model (E3) is shown. Hovering over the radar plot displays the numerical values of the probabilities (E4). At the bottom right, the \textbf{Stamp explorer} panel shows the science (F1), reference (F2) and difference (F3) image stamps associated with any point in the light curve, which can be downloaded for further analysis (F4), or displayed in full screen mode (F5). Users can switch between the previous or next stamps in time (F6), or select any particular date (F7) of the light curve which is contained in the public ZTF stream. Users can select between displaying cross hairs (F8) or simultaneously hovering and zooming (F9) over the stamps. They can also see the full alert information in the associated alert packet (F10). Note that those points in the light curve that do not pass the ZTF's real/bogus test will not have stamps available for display since they do not trigger an alert in the public stream. \label{fig:explorer_2}}
\end{figure*}

\begin{figure} [htb]
\centering
\includegraphics[width=0.4\textwidth]{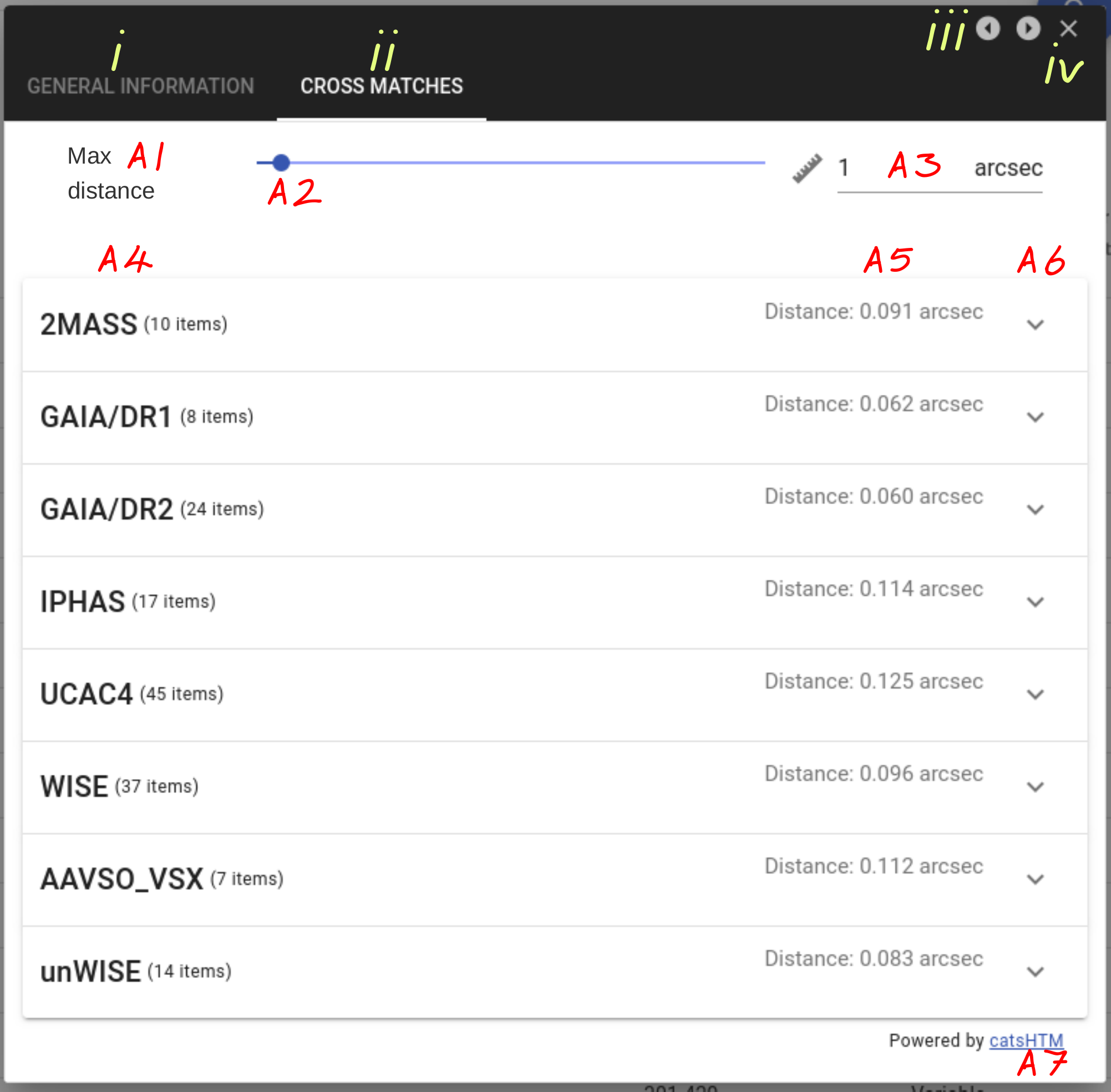}
\caption{The Object Cross matches view of the ALeRCE explorer. Labels \textit{i, ii, iii} and \textit{iv} as in Figure~\ref{fig:explorer_2}. This panel allows users to find the closest cross--matching sources in the catsHTM dataset, given a maximum cross--matching distance (A1) defined via a sliding bar (A2) or directly via its numeric value (A3). The closest cross--matches among different catalogs (A4) are shown with their associated distances (A5), allowing for an expanded view of the columns available in each catalog (A6). For more information, see the catsHTM (A7) reference \citep{2018PASP..130g5002S}. \label{fig:explorer_3}}
\end{figure}

\paragraph{ALeRCE Explorer (\url{http://alerce.online})} 

The ALeRCE explorer is the main tool to explore the astronomical objects recovered from the ZTF alert stream. Its landing page consists of two main sections: the Search and Results sections (see Figure~\ref{fig:explorer_1}). The Search section is where users can filter objects by selecting their unique identifier, or by selecting different combinations of classifier, class, class probability, number of detections, and sky coordinates. The Results section is where the results of the filtered objects are shown, sorted by classification probability or other variables. Clicking on an individual object will take the user to the object view page (see Figure~\ref{fig:explorer_2}). 

The object view page is divided into two tabs: the General Information and the Cross Matches tabs, with different panels each (see Figure~\ref{fig:explorer_2}). In the General Information tab users can see some basic statistics about the object, generate a finding chart, query different catalogs at the position of the object (NED, Simbad, TNS, PanSTARRS, or SDSS), or quickly see basic TNS information about the object. The user can see the object's light curve, including detections and non--detections, with the capability of plotting the raw difference light curve, a corrected apparent magnitude (which includes the contribution of the reference image), or a folded version of the corrected apparent magnitude using the best--fitting period. The light curve information can be downloaded as comma separated values (CSV), and every point in the light curve can be hovered over to see more information, or clicked on to show its associated image stamp. HiPS images and catalogs around the position of the object are shown using Aladin, with superimposed NED and Simbad clickable objects. The science, reference and difference image stamps associated with any point in the light curve can be shown in the Stamps section, where the stamps can be explored by selecting different dates or hovering over them, seen in full screen, or downloaded as fits files. The full Avro packet information can also be explored. The classification probabilities are shown in the Stamp and Light Curve Classifier tabs, where a radar plot is used to show the class probabilities assigned by the light curve or stamp based classifiers, if available. Finally, in the Cross Matches tab users can see all the cross--matches contained in the catsHTM set of catalogs for a given separation, which can be selected manually with a sliding bar (see Figure~\ref{fig:explorer_3}).

The ALeRCE explorer is where most of our web development has been focused, including new tools as requested by our community of users, but also new sources of data which in the future will allow for the multi--stream exploration of astrophysical objects. We are developing a modular data exploration library which will be gradually expanded to include new sources of streaming data\footnote{https://vue-components.alerce.online/}.

\begin{figure*} [htb]
\centering
\includegraphics[width=1.0\textwidth]{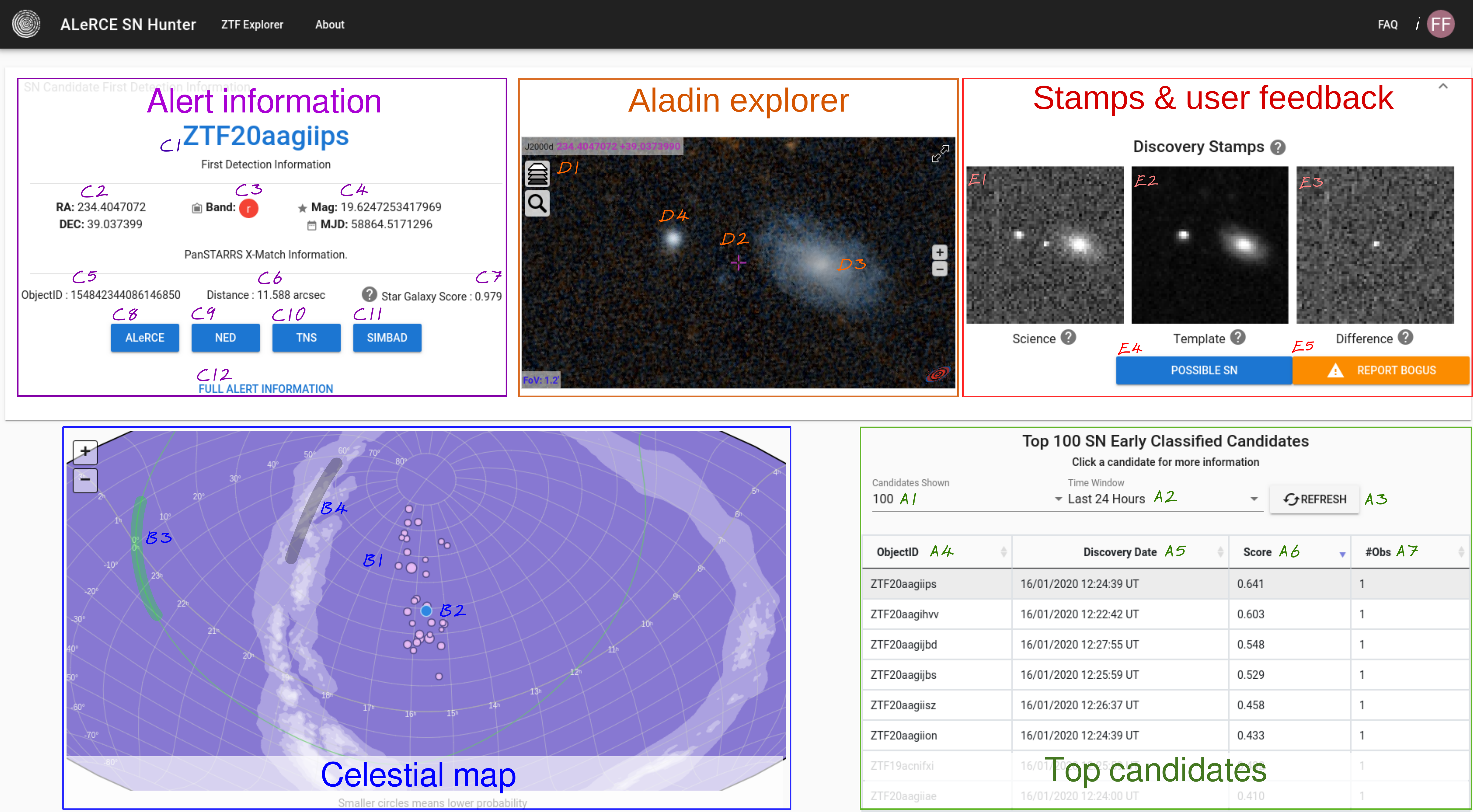}
\caption{The SN Hunter web interface (\url{http://snhunter.alerce.online}), which allows users to find the highest stamp--classification probability and most recent SN candidates in the ZTF alert stream in real--time. This tool is divided into five panels and is used by our collaboration to select candidates for submission to TNS. Starting at the bottom right, the \textbf{Top candidates} panel shows a list of the top 10 -- 1000 (default 100; A1) SN candidates in terms of their stamp classifier SN probabilities within the last 1-7 days (default 24 hours; A2). This list can be refreshed at any moment (A3). The results are shown in a paginated table sorted by either object identifier (A4), discovery date (A5), score or stamp classifier SN probability (A6), or number of detections (A7). Each candidate can be clicked on for exploration, opening up the top panels. At the bottom left, the \textbf{Celestial map} panel shows the spatial distribution of all the candidates in the Top candidates panel, with a circle size proportional to their score (B1) and centered around the currently selected candidate (B2). Also shown are the position of the ecliptic (B3) and Milky Way plane, where the white contour levels denote crudely the density distribution of Galactic stars (B4). At the top left, the  \textbf{Alert information} panel shows the information about the currently selected candidate, including its object identifier (C1); coordinates (C2); band (C3), magnitude and time (C4) at first detection; information about the closest PanSTARRS source, including its identifier (C5), distance (C6), and star galaxy score (C7, varying between 0 and 1 between  galaxies and stars). Links to the ALeRCE explorer Object view (C8), NED (C9), TNS (C10), and Simbad (C11) are provided. All additional information contained in the alert is also available for exploration (C12). At the middle top, the \textbf{Aladin explorer} panel provides an interactive Aladin window (D1) centered around the selected candidate (D2), where a host galaxy may be seen in PanSTARRS DR1 $gri$ color images (D3). Note that although there is a clear host galaxy associated with this candidate, its closest source is a star (D4), which explains the star galaxy score displayed in C7. Finally, at the top right, the \textbf{Stamps \& user feedback} panel is where the science (E1), reference (E2) and difference (E3) ZTF image stamps are displayed for the currently selected candidate. If users are logged in using a Google account (i), they can label candidates as possible SNe (E4) or report them as bogus (E5) in order to improve the stamp classifier training set. \label{fig:snhunter}}
\end{figure*}

\paragraph{SN Hunter (\url{https://snhunter.alerce.online})\label{sec;}} 

The SN Hunter platform allows users to visualize and explore the best and most recent SN candidates (see Figure~\ref{fig:snhunter}). These candidates are obtained using the convolutional neural network which powers the ALeRCE stamp classifier and can be seen in the SN Hunter just seconds after being received from ZTF. Users can see the spatial distribution of the candidates in celestial coordinates and in comparison to the Milky Way plane or the ecliptic, as well as a table which shows them sorted by classification probability, discovery date, or number of observations. Selecting a candidate displays an Aladin HiPS image at the location of the object, as well as the science, reference and difference images contained in the Avro file. The candidates's unique identifier, coordinates, first observation properties, and the properties of the closest PanSTARRS object are also shown, as well as links to the ALeRCE explorer for the same object, or for NED, TNS and Simbad sources around the position of the object. Users can also see the full alert information contained in the original Avro file of the alert by clicking in Full Alert Information button. 

A key feature of the SN Hunter is the ability to receive feedback from users who have logged in. If a candidate appears to be bogus, users can label the candidate as such to further enhance the training set. Moreover, if the candidate appears to be a SN or extragalactic transient, the user can label it as a possible SN to be sent to the ALeRCE reporter tool (see below). The list of possible SNe can then be explored by the team with our reporter tool, which can then be used to submit targets to the TOMs for follow--up.

\begin{figure*} [htb]
\centering
\includegraphics[width=0.8\textwidth]{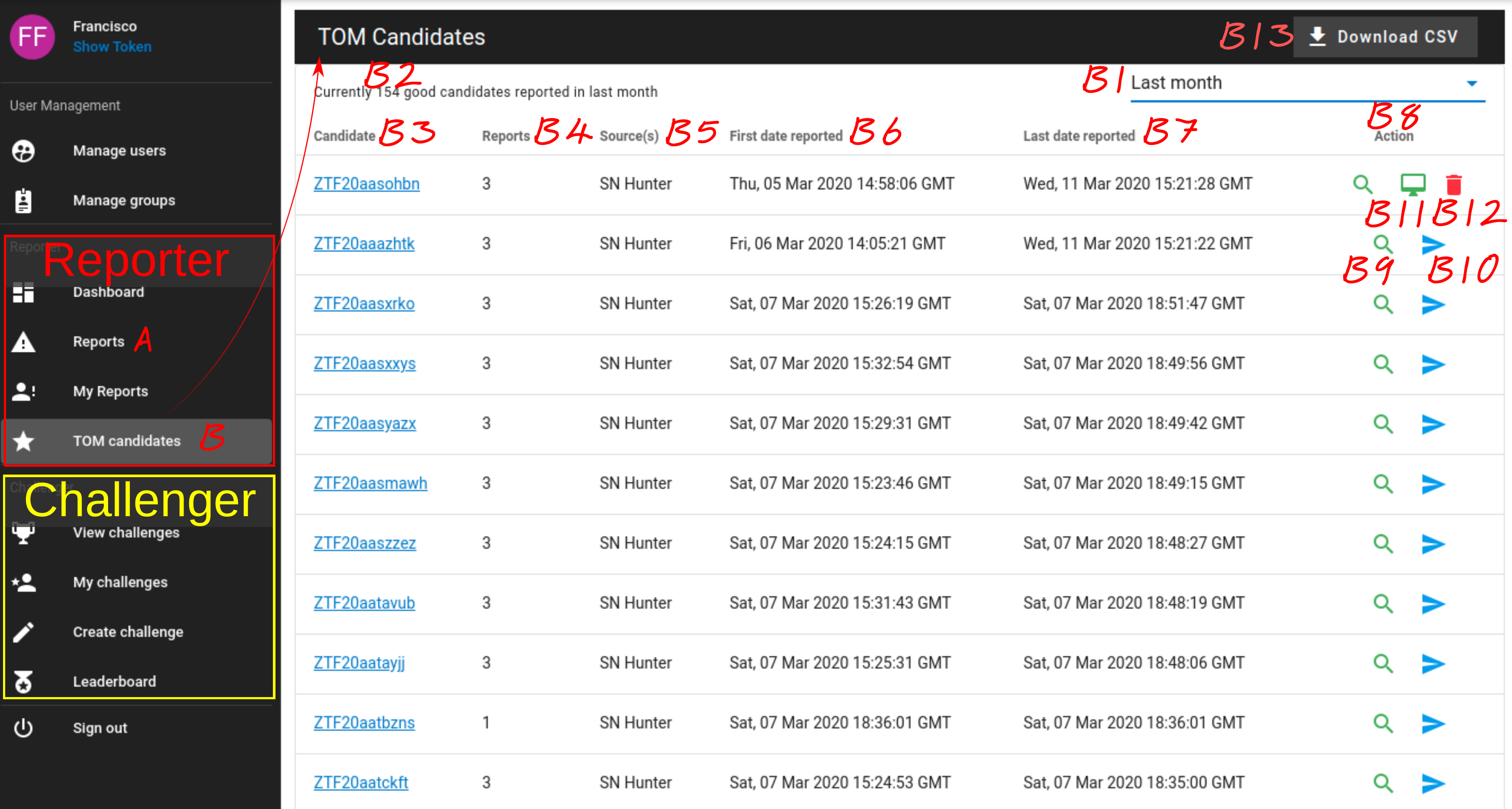}
\caption{The ALeRCE reporter web interface (\url{http://reporter.alerce.online}) is used to manage user input in the ALeRCE ecosystem. Here we show two types of inputs: the Reporter tool, that manages input labels from the SN hunter, either bogus (A) or possible SN (B), which in the latter case become candidates to be sent to the TOM; and the Challenger tool, which we use to manage data classification challenges or hackathons. In the TOM list of possible SNe, users can select a given period of time of recently reported candidates (B1), which returns a given number of candidates (B2). Users can explore the object identifiers (B3), number of independent reports for the given candidate (B4), source of the label (B5), date of first (B6) and last (B7) reports, and possible actions (B8). Among the possible actions, users can explore who has reported a candidate (B9), create a target for observations in the TOM Toolkit (B10), edit the observational properties of an already created TOM candidate (B11), or remove the target from the TOM Toolkit (B12). The full list can be downloaded as a CSV file (B13).  \label{fig:reporter}}
\end{figure*}

\paragraph{Reporter (\url{https://reporter.alerce.online})}

The ALeRCE reporter tool is a platform which serves to manage user feedback in general (see Figure~\ref{fig:reporter}). As of Jun 2020 it serves three purposes: to manage the feedback provided by the SN Hunter interface, to connect with the TOM Toolkit interface, and to manage internal data classification challenges. The user feedback provided via the SN Hunter consists of bogus alert labels, for those alerts which appear to be bogus; and possible SN alert labels, for those alerts or groups of alerts which appear to be originated by extragalactic transients. The connection of SN candidates with the TOM Toolkit interface is also done from the reporter tool, sending users to the TOM Toolkit Interface after clicking on a reported candidate. Finally, the reporter tool can be used to create data challenges, manage associated user entries, produce metrics and confusion matrices, and show leader boards as in \href{https://www.kaggle.com/}{Kaggle}. The data challenges are key for the collaboration's periodic hackathons, where we set different classification challenges and which motivate the ML team to develop new ideas and tools.

\paragraph{TOM Toolkit Plugin (\url{https://tom.alerce.online})}

This platform is used to manage and submit candidates to the TOM Toolkit (\url{https://lco.global/tomtoolkit/}). Users that have access rights to the ALeRCE reporter can connect with the TOM Toolkit via this interface, allowing them to submit observational requests with detailed instrumental specifications to the queue of different observatories.

\begin{figure*} [htb]
\centering
\includegraphics[width=0.8\textwidth]{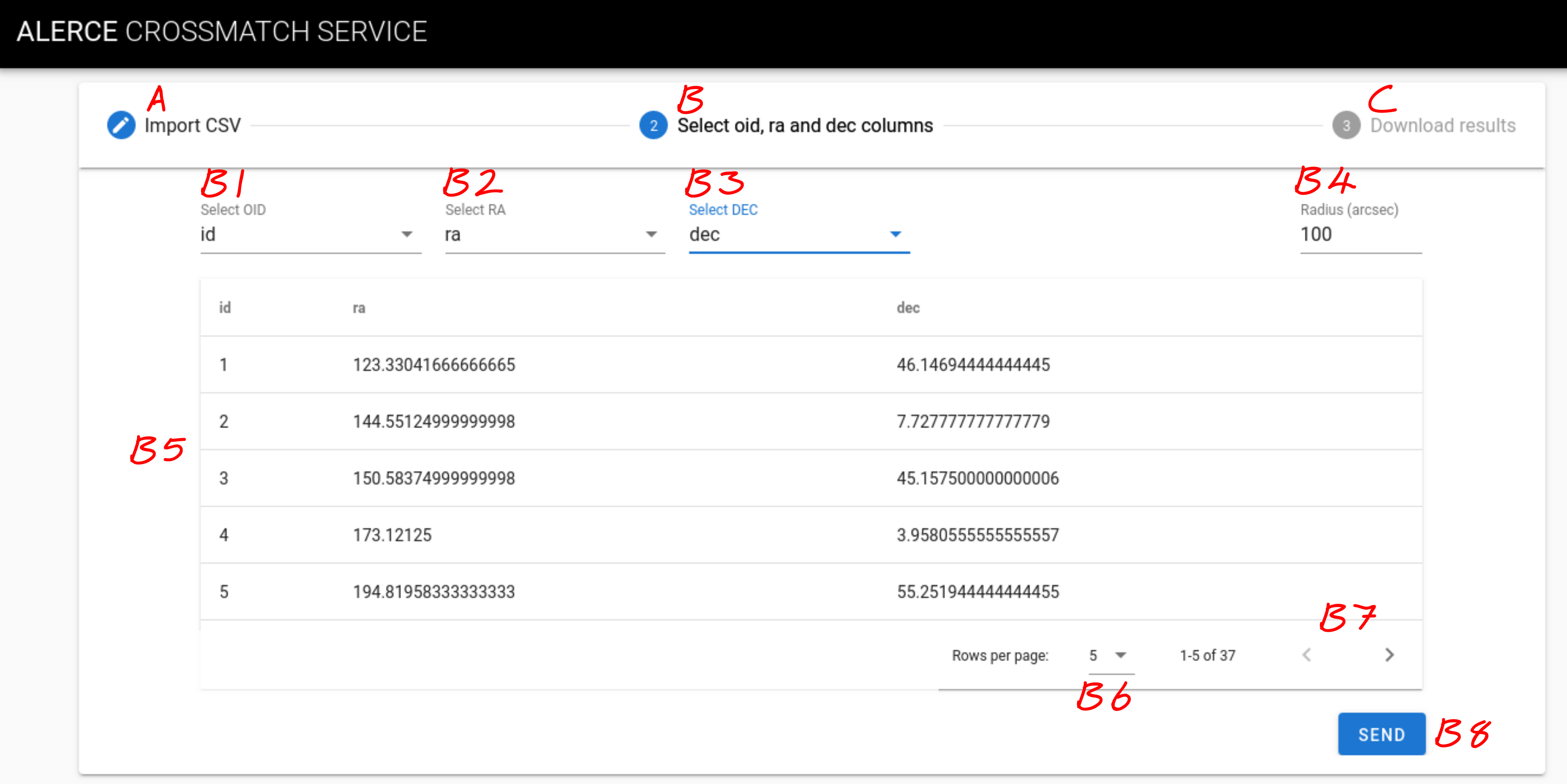}
\caption{The cross--match service interface (\url{https://xmatch.alerce.online}). Users can input arbitrary catalogs as csv files to be cross--matched to the ZTF database. The procedure consists in selecting an input catalog CSV file (A), and then indicating the columns in the file which will be used as identifier (B1), right ascension (B2) and declination (B3), as well as the maximum radius used to search for the closest cross--matching source (B4). The information provided allows for the partial exploration of the input file (B5) by a given number of rows (B6) in paginated form (B7). After submitting the catalog (B8), users can visually explore and download the cross--matched catalog (C). \label{fig:xmatch}}
\end{figure*}

\paragraph{Xmatch Service (\url{http://xmatch.alerce.online})} \label{sec:xmatch}

ALeRCE provides a cross--match service which allows users to submit an arbitrary CSV file with objects and coordinates of their favorite targets (see Figure~\ref{fig:xmatch}). After a file is uploaded, the user is asked to select the names of the identifier, right ascension and declination columns. After this is done the closest objects in ZTF are returned, adding several columns from the ALeRCE object table to the submitted objects. A paginated table is shown for exploration, and the output can be downloaded as a CSV file.

\subsubsection{APIs} \label{sec:API}

All the interactions between the Web Interfaces and the database or the Avro/stamp repository are done via APIs. These APIs serve most of ALeRCE's data exploration tools following the principle of maximizing the modularization of our different services. They are also the key elements which will allow ALeRCE to integrate seamlessly with the astronomical time--domain ecosystem. These APIs are documented in the ALeRCE API Documentation webpage: \url{https://alerceapi.readthedocs.io/en/latest/}. Here we describe the services available as of Jun 2020:

\paragraph{ZTF Database Access Service (\url{http://ztf.alerce.online})} 

This service allows users to query the ALeRCE database tables without needing any authentication. This API includes services to query objects filtered by unique object identifier, number of detections, class, class probabilities, coordinates, or detection times. Users can also get the associated SQL command for a given query, all the detections for a given object, all the non--detections for a given object, the classification probabilities for a given object, or the features used as input for the ML classifiers for a given object. The documentation can be found in \url{https://alerceapi.readthedocs.io/en/latest/ztf_db.html}. This service is used in the ALeRCE explorer and the SN Hunter (see Section~\ref{sec:web}).

\paragraph{Avro/Stamps Service (\url{http://avro.alerce.online})}

This service allows users to access the alert Avro files and their associated stamps. The input is the unique object identifier and the unique stamp identifier. Users can get the Avro file, a specific field from an Avro file, or the science, reference and difference image stamps contained in an Avro file. The documentation can be found in \url{https://alerceapi.readthedocs.io/en/latest/avro.html}. This service is used in the ALeRCE explorer and the SN Hunter (see Section~\ref{sec:web}).

\paragraph{ZTF Xmatch Service (\url{http://xmatch-api.alerce.online})}

This service allows users to submit an arbitrary catalog and get the nearest ZTF sources, their separation, and their properties. It is used in the Xmatch interface (see Section~\ref{sec:web}).

\paragraph{catsHTM Crossmatch Service (\url{http://catshtm.alerce.online})}

This service allows users to do cone searches to a given location using the catsHTM catalogs \citep{2018PASP..130g5002S}. This includes cone searches returning all the objects closer than a given distance from all the catalogs, from a specific catalog, or only the closest object from all or a given catalog. This service is used in the ALeRCE explorer Cross Matches view (see Section~\ref{sec:web}). The documentation, indicating also a list of all the available catalogs, can be found in \url{https://alerceapi.readthedocs.io/en/latest/catshtm.html}.

\paragraph{TNS Crossmatch Service (\url{http://tns.alerce.online})}

This service allows users to query TNS information about an object centered around a given position in the sky. It queries the TNS API and returns the TNS name, type and redshift, and it is used by the ALeRCE explorer General Information Tab (see Section~\ref{sec:web}).

\begin{figure} [htb]
\centering
\includegraphics[width=0.45\textwidth]{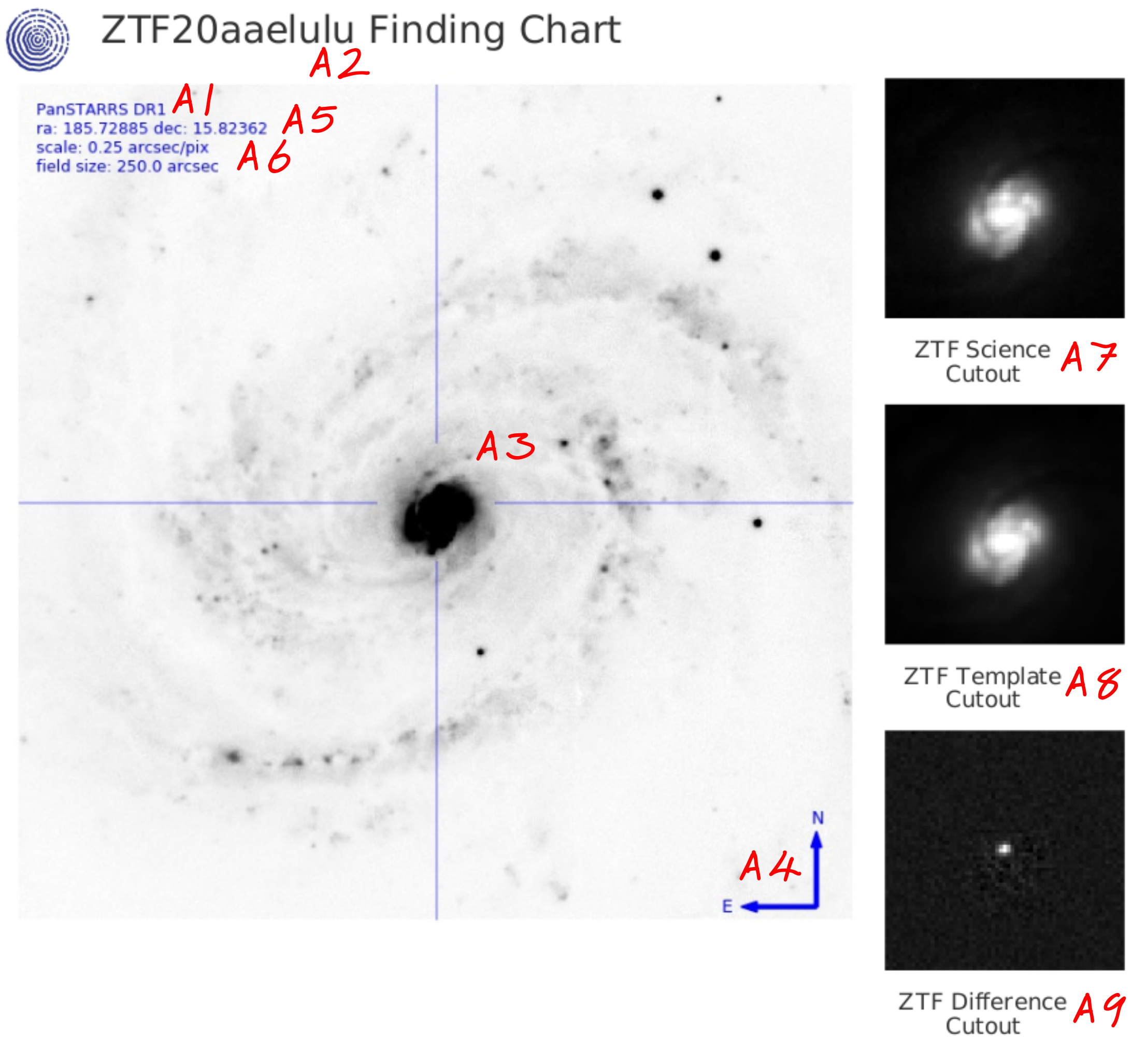}
\caption{\label{fig:findingchart} A section of the finding chart generated automatically for object ZTF20aaelulu, or SN 2020oi, a Type Ic SN that occurred in the nearby galaxy M100. The finding chart shows a PanSTARRS DR1 image (A1) centered around this object (A2, A3), indicating the direction of the north and east axes (A4), the coordinates (A5), and the pixel scale and field size (A6). It also shows the ZTF science (A7), reference (A8) and difference image stamps (A9). Additional information, such as the coordinates in a different format, magnitude statistics, or the time of first and last detection, are also included. Note that this SN was reported to TNS by ALeRCE after being classified as a possible SN with just a single detection using the SN Hunter tool (see Figure~\ref{fig:snhunter}).}
\end{figure}

\paragraph{Finding Chart Service (\url{http://findingchart.alerce.online})}

This service provides a finding chart associated with a given object's unique identifier. It returns a pdf file with a PanSTARRS reference image indicating the location of the candidate, as well as the science, reference and difference image stamps. An example finding chart can be seen in Figure~\ref{fig:findingchart}. This service is used in the ALeRCE explorer (see Section~\ref{sec:web}).

\paragraph{Python API Client}

We provide a Python client for easier access to the previous API services. It can be installed via \verb+pip+ and is documented in \url{https://alerce.readthedocs.io/en/latest/}. You can find examples of how to use the client in the \href{https://github.com/alercebroker/usecases}{use case notebooks}.

\section{Results} \label{sec:results}

\begin{figure} [htb] 
\centering
\includegraphics[width=0.5\textwidth]{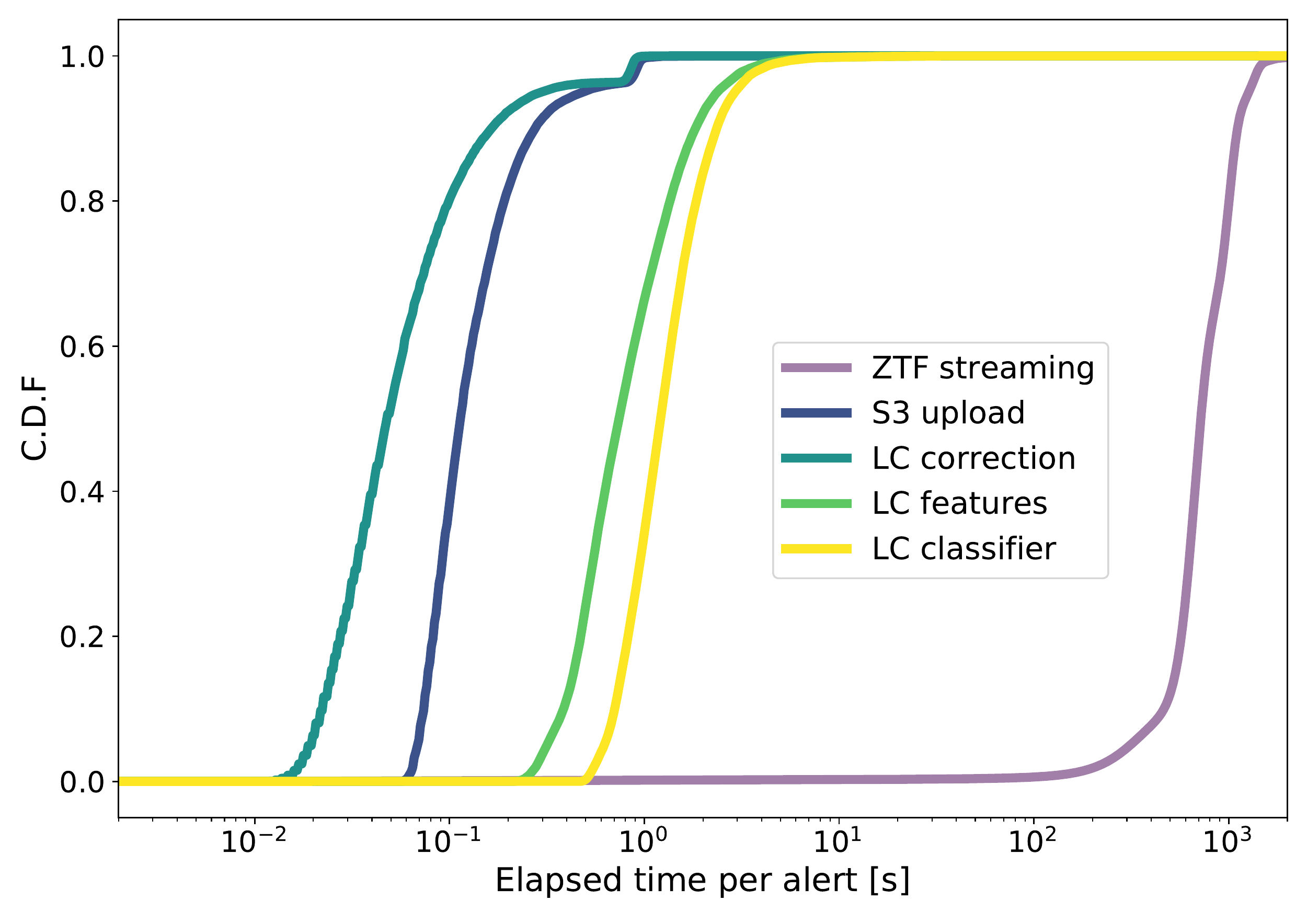}
\caption{Cumulative distribution function (CDF) of ZTF streaming times compared to the CDF of ALeRCE pipeline processing times. The ZTF streaming times corresponds to the difference between the reported observation time and the alert ingestion time, obtained empirically in a typical night of operations. The ALeRCE pipeline step elapsed times stands for the time needed for an alert to move from ingestion to the completion of a given step, including CPU and wait times.  In this figure we consider an incoming alert rate of about 25 s$^{-1}$ (c.f., we expect about 5 and 350 s$^{-1}$ for ZTF and LSST on average, respectively). The \emph{embarrassingly parallel} nature of the processing steps suggests that our infrastructure should scale linearly with the number of incoming alerts to manage the LSST alert stream. \label{fig:exectimes}}
\end{figure}

The ALeRCE broker has processed \nalerts alerts from the public ZTF stream, at a rate of about 5$\times 10^7$ per year, which corresponds to about 1.4$\times 10^5$ per night, or about 5 alerts per second on average. This is $\sim 80 \times$ less than the expected alert rate of LSST of about $10^7$ per night. However, the ZTF public stream alert production rate is not constant, with some nights producing a few million alerts, which we have been able to ingest without significant wait time increases. In Figure~\ref{fig:exectimes} we show the distribution of processing times (CPU + waiting times) at the different steps of our pipeline for a typical ZTF night, including the distribution of ZTF streaming times (time between observation and ingestion) for comparison. With our current infrastructure we can process ZTF alerts in real--time, with classification delays being dominated by the ZTF streaming times. The latest version of the ALeRCE pipeline has been tested at rates of about 150 alerts per second, which is approximately 45\% of the expected rates of LSST. 

As of Jun 2020, we have \nobjects objects, \nalerts detections, and \nnondet non--detections in our database. There are \nLCclf objects classified by the light curve classifier and \nstampclf objects classified by the stamp classifier, which started being applied to new alerts in Aug 2019. For a distribution of the ML inferred classes in these samples, see our accompanying papers \citep{earlyclassifier,lateclassifier}. The associated confusion matrices can be seen in Figures~\ref{fig:lateconfusion} and \ref{fig:earlyconfusion} and a comparison between the two classifiers can be seen in Figure~\ref{fig:stampvsLC}. Note that our classifiers are continuously improving and that the choice of model is not based solely on a balanced accuracy score, but also on a study of the relative frequency and spatial distribution of classes in the unlabeled set, which we have found to be an important verification when the training set is not representative of the unlabeled set.

\begin{figure} [htb!] 
\includegraphics[width=0.45\textwidth]{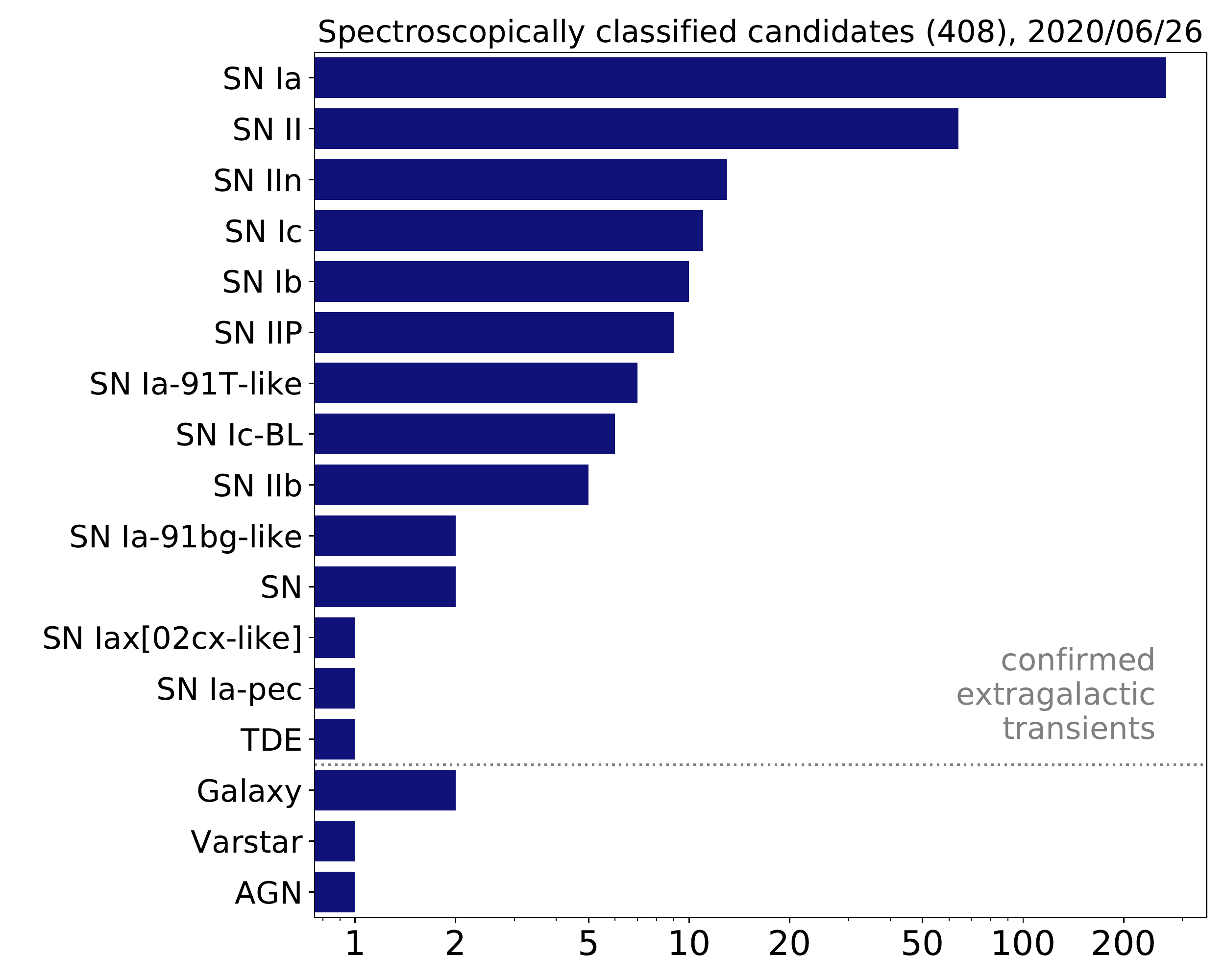}
\caption{The sample of spectroscopically classified transients first reported by ALeRCE to TNS, from \nsncand SN candidates submitted based on their first alert. Out of \nspec candidates observed spectroscopically, \nsnany are confirmed as SNe, \nunclear are unclear, \ngalsn is a likely SN misclassified as a galaxy, and \nnosn are not SNe. Of the \nsnany confirmed SNe, \nsnia are SNe Ia, \nsnii are SNe II, \nsnibc are SNe Ib/c, \nsnpec are other peculiar types, and \nsnnoclass are classified just as SNe. The \nunclear unclear cases, both of which had SN--like light curves, are AT 2019yzs (\href{http://alerce.online/object/ZTF19adcbnty}{ZTF19adcbnty}), which could be a SN, TDE, or AGN; and AT 2020bdh (\href{https://alerce.online/object/ZTF20aaivtof}{ZTF20aaivtof}), which has a very noisy spectrum. The likely galaxy misclassification is AT 2019tkd (\href{https://alerce.online/object/ZTF19aciiuta}{ZTF19aciiuta}), which also has a SN--like light curve. The \nnosn cases confirmed as not SNe are AT 2019qiz (\href{https://alerce.online/object/ZTF19abzrhgq}{ZTF19abzrhgq}), which is a TDE;	AT 2020fx (\href{https://alerce.online/object/ZTF20aadymod}{ZTF20aadymod}), which is a high proper motion star in the line of sight of a galaxy; AT 2019uzg (\href{https://alerce.online/object/ZTF19acssnul}{ZTF19acssnul}), which is a badly subtracted galaxy, likely a bad zero point calibration; and AT2020csk (\href{https://alerce.online/object/ZTF20aaodhzr}{ZTF20aaodhzr}), which is an AGN. \label{fig:totalsne}}
\end{figure}

\begin{figure} [htb!] 
\centering
\includegraphics[width=0.5\textwidth]{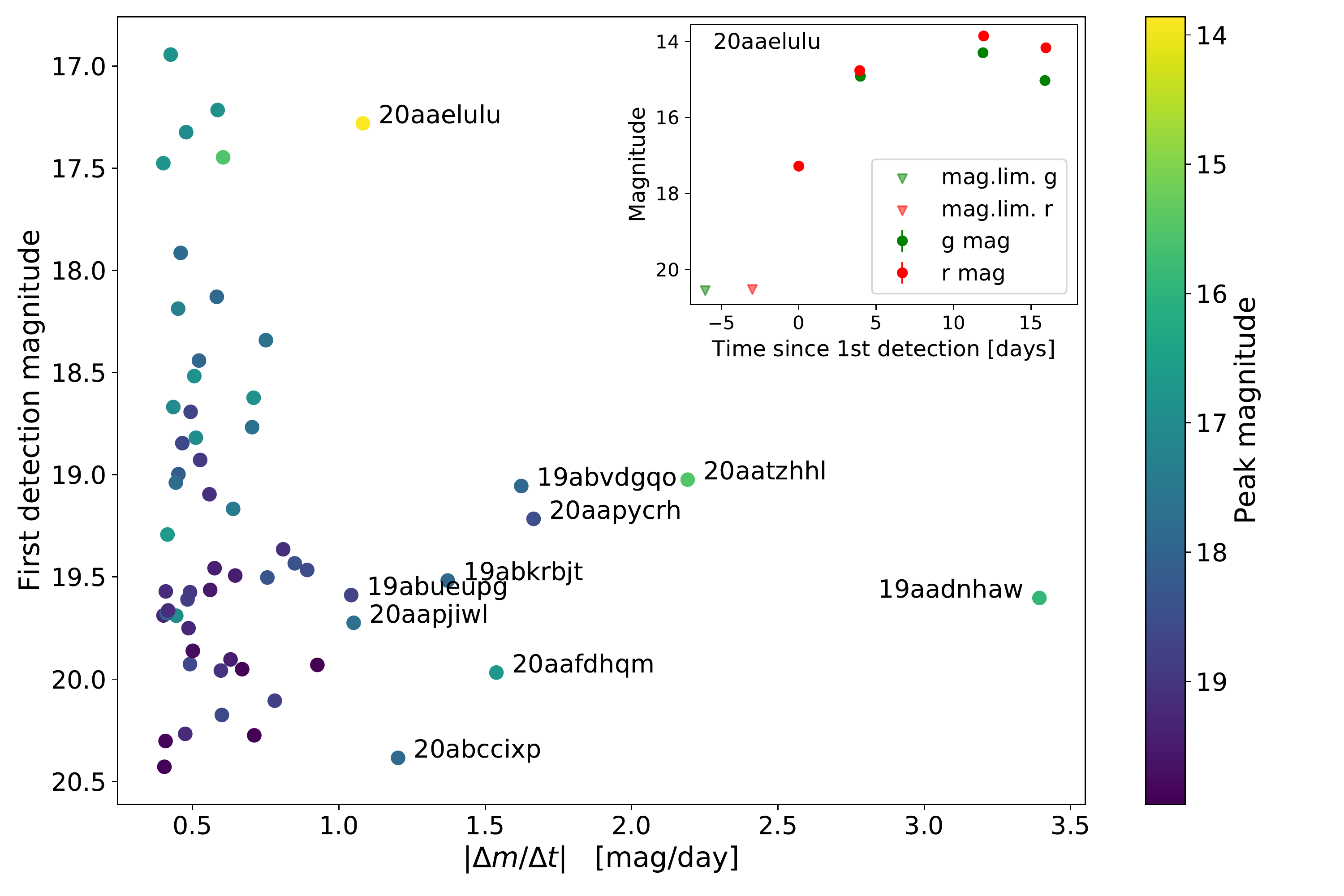}
\caption{Detection magnitude vs. magnitude rise rate at time of detection for the SN candidates reported to TNS by ALeRCE based on their first alert image stamps. The color indicates the peak magnitude of the candidate. We only show candidates detected rising faster than 0.4 mag/day, a sample which includes \nsnrisegtptfour SN candidates. We individually label \nsnrisegtone candidates which rose faster than one mag/day at detection. Of these candidates, \href{https://alerce.online/object/ZTF19abueupg}{ZTF19abueupg}, \href{https://alerce.online/object/ZTF20aapjiwl}{ZTF20aapjiwl}, \href{https://alerce.online/object/ZTF20aapycrh}{ZTF20aapycrh}, \href{https://alerce.online/object/ZTF20aatzhhl}{ZTF20aatzhhl} and \href{https://alerce.online/object/ZTF20abccixp}{ZTF20abccixp} are SNe II; \href{https://alerce.online/object/ZTF20aaelulu}{ZTF20aaelulu} is a SN Ic (shown in the inset plot); \href{https://alerce.online/object/ZTF19abvdgqo}{ZTF19abvdgqo} is a SN Ib; \href{https://alerce.online/object/ZTF19abkrbjt}{ZTF19abkrbjt} is a SNe Ia; \href{https://alerce.online/object/ZTF20aafdhqm}{ZTF20aafdhqm} is a transient which coincided with a previous SN candidate (PS1-13dgc); and \href{https://alerce.online/object/ZTF19aadnhaw}{ZTF19aadnhaw} is probably a nova based on the shape of its light curve and the presence of a blue stellar source at its position. \label{fig:fastrisers}}
\end{figure}

An important tool to connect ALeRCE with the SN community of users is the SN Hunter. We have used it to report \nsncand previously unreported astrophysical transient candidates to TNS, \nspec of which have been classified spectroscopically  (with 1\% contamination among those classified spectroscopically, see Figure~\ref{fig:totalsne}). Among these, we have found \nsnrisegtptfour SN candidates rising faster than 0.4 mag/day, and \nsnrisegtone faster than 1.0 mag/day, at discovery (see Figure~\ref{fig:fastrisers}). In the process, we have visually inspected about 20,000 candidates, saving in our database more than 6500 bogus candidates since Oct 2019 and 1100 transient candidates since Jan 2020, when we added the \emph{Bogus} and \emph{Possible SN} buttons to the SN Hunter, respectively. The bogus examples have been used to increase the size and diversity of our training set and have resulted in significant improvements to the stamp classifier.

\begin{figure} [htb!] 
\centering
\includegraphics[width=0.5\textwidth]{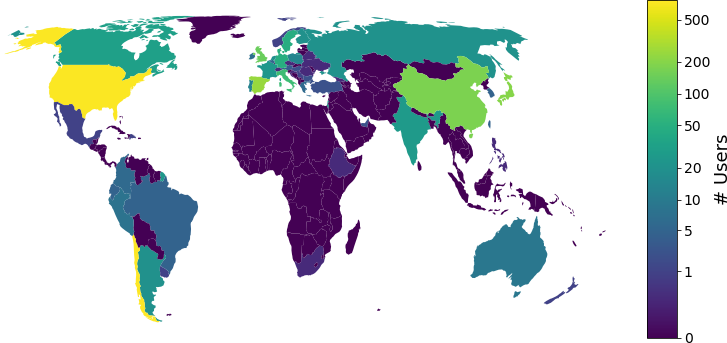}
\caption{The geographic distribution of users of the ALeRCE Explorer according to Google Analytics. The number of users is estimated counting the unique combinations of device \& browser accessing our website. In total, we have more than \nusers estimated users coming from \ncountries different countries accessing the ALeRCE Explorer. \label{fig:countries}}
\end{figure}

We are slowly building an international community of users. In order to facilitate the adoption of our tools by the community, we do not require users to create accounts to access our system, which makes it difficult to precisely estimate the number of ALeRCE users. However, we can use Google Analytics\footnote{https://analytics.google.com} to quantify our online community of users. Since Jul 2019, when Google Analytics was added to the ALeRCE Explorer and SN Hunter tools, we have had 2.1/1.3 k users (unique combinations of device and browser, as per the Google definition) and 7.7/2.2 k sessions in the AleRCE Explorer/SN Hunter. This does not include the use of APIs or direct connections to our database. Our users are currently distributed in \ncountries countries (see Figure~\ref{fig:countries}), with the top ones being Chile (27.2\%), U.S. (25.8\%), Spain (8.9\%), Japan (7.3\%), China (6.5\%), and U.K. (5.1\%).  We are continuously listening to our users to include new features and we have created new use case jupyter notebooks for different science cases. We encourage users to create additional use case notebooks and contribute to our open source repository (\url{https://github.com/alercebroker/usecases}).

\section{Discussion and conclusions} \label{sec:conclusions}

The ALeRCE broker is a new--generation astronomical alert broker, processing alerts in real--time from ZTF and preparing to become a community broker for LSST. We are an interdisciplinary, inter--institutional and international team led from Chile, using Agile methodologies to develop new digital components for the astronomical time--domain ecosystem in the era of large etendue telescopes. 

In this document we have reported the motivation, challenges, methodologies and first results of the ALeRCE broker. The main motivation for ALeRCE is to provide a rapid classification of events to enable fast follow--up and characterization, but also to provide a systematic classification of all variable objects for a self--consistent analysis of large volumes of events in the observable Universe. Our primary scientific drivers are the study of transients, variable stars, and AGN, but we also provide Solar System object classifications for further analysis.

We describe the infrastructure, processing steps, data products, tools \& services that work in real--time. We ingest, aggregate, and cross--match the alert stream, and apply two ML based classifiers to the data (see Section~\ref{sec:classifier}). First, a stamp classifier is applied to all alerts associated with previously unreported objects using the first image stamps as input and a simple taxonomy. Second, a light curve classifier with a more complex taxonomy is applied to all objects with $\geq6$ detections in $g$ or $\geq6$ detections in $r$. We are also experimentally applying outlier detection methods to the data, which we hope to make public in real--time after significant testing is done. To our knowledge, ALeRCE was the first public broker to provide real--time classification of the ZTF alert stream into an astrophysically motivated taxonomy based on the alert image stamps or their light curves.

Regarding the processing of the data, our processing times per alert are of the order of seconds, significantly smaller than the current ZTF streaming times (see Section~\ref{sec:results}). Moreover, we have run experiments at ingestion rates similar to those expected for LSST. 

Our database contains object, detection and non--detection based families of tables, with increasing numbers of rows, which are indexed for fast query speeds. All relevant tables are public with read--only access, although we recommend accessing them via our different APIs which power all our web--based services and Python client. We provide extensive documentation for our different data products and services, which can be found in our main website, \url{http://alerce.science}. All our data products, documentation, tools and services are summarized in Table~\ref{tab:dataproductsservices}.

Apart from providing a classified stream of data upon request, our two most important web services are the ALeRCE Explorer (\url{https://alerce.online}) and the SN Hunter (\url{https://snhunter.alerce.online}), which are publicly available and described in detail in Sections~\ref{sec:web}. The ALeRCE Explorer is the main tool to explore the objects contained in the ZTF public stream, allowing for simple queries and providing a user friendly visualization of their light curves, cross--matches, image stamps and classification probabilities. The SN Hunter tool is targeted for the transient community to enable a rapid reaction, allowing users to quickly explore and provide feedback on the latest SN candidates contained in the stream. We use this tool to submit new SN candidates to the TNS at an average rate of about 9 per night, with \nsncand reported candidates since Aug 2019. We also use this tool to select candidates for follow--up via the TOM Toolkit.

An important goal of ALeRCE is to provide a good user experience, which should allow for a smooth transition into a time--domain ecosystem dominated by large alert streams and automated components where astronomers and data scientists are not replaced, but instead are aided by ML tools to achieve new discoveries. Thus, we are developing different modular components for the visualization of the alert stream data, optimized for usability after testing with our community of users in regular tutorials and hackathons. The use of Agile methodologies with a fully dedicated interdisciplinary team of engineers and astronomers has been critical to develop ALeRCE at the speed required by the community. Collaboration remains essential among brokers to bring a more diverse set of ideas into our community and add resilience to the time--domain ecosystem in the era of large etendue telescopes. 

One of the biggest challenges ahead for ALeRCE is the ability to scale to significantly larger streams, from $\sim 1.4 \times 10^5$ alerts per night to $> 10^7$ alerts per night; and with significantly more objects generating alerts, from a few $10^7$ objects to $> 10^9$ objects. For this, we will need to migrate some of our tables from a SQL, centralized database engine, to a NoSQL, distributed database engine (e.g., Cassandra, MongoDB). We are running different tests to determine the efficiency and cost of the different available solutions in collaboration with other brokers (Fink). Another important challenge is to determine what fraction of our storage and computing services should be located in the cloud (e.g., AWS, where we currently operate some of our services) vs on--premise infrastructure. It seems likely that the answer will be a hybrid solution, with cloud and on--premise infrastructure optimized for a better user experience while minimizing the operational costs.

Achieving more complex taxonomies in an era of multi--stream, multi--messenger astronomy is another important challenge ahead. In fact, the large number of events expected, combined with the addition of heterogeneous streams spanning different depths, cadences, wavelengths, and messengers will likely unveil new populations which would not have been possible to identify otherwise.  Encompassing the full diversity of variable classes in the Universe with a fixed taxonomy is unfeasible, and thus our taxonomy will continue to grow and evolve with time. Eventually, a combination between domain knowledge via supervised training, with unsupervised, more data--driven taxonomies, will become necessary. Training and classifying with missing data, as most streams of data will be sparse in comparison to that of LSST, will also become important.

Regarding the challenges of ML classification, we are trying different strategies. We are introducing new features, e.g. a complex number extension to the IAR model that allows for positive as well as negative autocorrelation \citep[CIAR,][]{2019A&A...627A.120E}, further expanded to bivariate or higher dimensional time series and to include different covariance structures. From these models we expect to extract useful features for classification, as well as be able to do prediction, interpolation and forecasting on time series. We are also testing ways to combine real, augmented and simulated data; new ways to combine and expand our Stamp and Light Curve classifiers; or different recurrent neural networks applied to the light curve \cite[e.g.,][]{2019PASP..131k8002M} and images stamp series \cite[e.g.,][]{2019PASP..131j8006C}; or different outlier detection methods.

Finally, we note that, given the continuously evolving nature of ALeRCE, this document provides a snapshot of the current status of ALeRCE as of Jun 2020. We are constantly listening to our community of users in an effort to introduce new data products, tools and services. Our preferred way of communication is through issues in our GitHub repositories (\url{https://www.github.com/alercebroker}), but users can also contact us directly via \url{https://alerce.science}.

\acknowledgments

The authors acknowledge support from
the Chilean Ministry of Economy, Development, and Tourism's Millennium Science Initiative through grant IC12009, awarded to the Millennium Institute of Astrophysics (FF, GCV, ECN, PAE, PSS, JA, FEB, RCD, MC, FE, SE, PH, GP, ER, IR, DR, DRM, CV, IAM, NA, JB, AC, DDC, CDO, RK, AM, WP, MPC, LSG, AS, CSC, JRV)
and from the National Agency for Research and Development (ANID) grants: 
BASAL Center of Mathematical Modelling AFB-170001 (FF, ECN, PAE, IR, DRM, CV, IAM, JCM, AM, LSG, JSM, CSC, EV) 
and Centro de Astrofísica y Tecnologías Afines AFB-170002 (FEB, MC, DDC, PSS); 
FONDECYT Regular \#1200710 (FF), \#1190818 (FEB), \#1200495 (FEB), \#1171273 (MC), \#1201793 (GP);  
FONDECYT Postdoctorado \#3200250 (PSS) and \#3200222 (DDC);
ANID infrastructure funds QUIMAL140003 and QUIMAL190012;
Magister Nacional 2019 \#22190947 (ER).
We acknowledge support from REUNA Chile who hosts and maintains some of our infrastructure.
This work has been possible thanks to the use of AWS-U.Chile-NLHPC credits.
Powered@NLHPC: This research was partially supported by the supercomputing
infrastructure of the NLHPC (ECM-02).

\software{Aladin \citep{2000A&AS..143...33B}, Apache ECharts\footnote{\url{https://echarts.apache.org}}, Apache Kafka\footnote{\url{https://kafka.apache.org/}}, Apache Spark \citep{zaharia2016apache}, ASTROIDE \citep{ASTROIDE}, Astropy \citep{2013A&A...558A..33A}, catsHTM \citep{2018PASP..130g5002S}, Dask \citep{matthew_rocklin-proc-scipy-2015}, FATS \citep{2017ascl.soft11017N}, Grafana\footnote{\url{https://grafana.com/}}, Imbalanced-learn \citep{JMLR:v18:16-365}, ipyladin \citep{2020ASPC..522..117B}, Jupyter \citep{Kluyver2016jupyter}, Keras \citep{chollet2015keras},
Matplotlib \citep{4160265}, NED \citep{2017AJ....153...37S}, P4J \citep{2012ITSP...60.5135H}, Pandas \citep{mckinney2010data}, Prometheus\footnote{\url{https://prometheus.io/}}, Python \citep{van1995python}, scikit-learn \citep{pedregosa2011scikit}, Simbad-CDS \citep{2000A&AS..143....9W}, Tensorflow \citep{abadi2016tensorflow}, Vue\footnote{\url{https://vuejs.org/}}, Vuetify\footnote{\url{https://vuetifyjs.com/}}, PostgreSQL\footnote{\url{https://www.postgresql.org/}}, XGBoost\footnote{\url{https://xgboost.readthedocs.io/}}.}

\appendix

\section{Light Curve Correction derivation} \label{sec:LCcorrection_appendix}

\subsection{Light Curve Fluxes}

An alert is originated when a significant flux is detected at some location of a difference image between a science and reference images. In the ZTF alert stream, the difference and reference fluxes are reported for every alert. The science flux is not reported, but it can be recovered  from the difference and reference images. The difference flux is reported by its absolute magnitude, $m_{\rm diff}$, and sign, $\rm sgn$; and the reference flux is reported by the PSF photometry magnitude, $m_{\rm ref},$ of the closest source in the reference, with associated errors, distance and shape parameters. This leads to three types of cases: 1) the closest source in the reference coincides with the location of the alert, and it is unresolved; 2) the closest source in the reference coincides with the location of the difference image alert, but it is resolved; and 3) the closest source does not coincide with the position of the  difference alert. In 1) the science flux can be recovered exactly, in 2) it can be recovered plus a constant which depends on how much contamination from an extended source occurs in the reference, and in 3) one needs to assume that the science flux is equal to the difference flux. These cases are typically represented by variable stars (1), AGNs (2), or transients (3).
Since it is not possible to know a priori which correction should be applied to each object, e.g., it is difficult to distinguish an AGN from a nuclear transient until the flux evolution can be observed, we report both the corrected photometry, which is useful for variable stars and AGNs, and the uncorrected photometry, which is useful for transients.

If the reference source is resolved, its reported flux contains two components: a variable/compact component, which is normally the object of study, and a static/extended component, which is difficult to separate using only the ZTF photometry. Because of the convolution done during the image difference process, the extended component should not contribute to the difference flux. Then, we note the following relations:
\begin{align}
    f_{\rm ref} &= f^{\rm ext}_{\rm ref} + f^{\rm var}_{\rm ref} \label{eq:frefcomp},\\
    f_{\rm sci} &= f^{\rm ext}_{\rm sci} + f^{\rm var}_{\rm sci}, \\
    {\rm sgn} f_{\rm diff} &= f^{\rm var}_{\rm sci} - f^{\rm var}_{\rm ref} \label{eq:sgnfdiff},
\end{align}
where $f_{\rm ref}$ is the reference flux, $f_{\rm sci}$ is the science flux, $\rm sgn$ is the sign and $f_{\rm diff}$ is the absolute value of the difference flux, $f^{\rm ext}_{\rm ref}$ is the contribution from the extended component in the reference image, $f^{\rm var}_{\rm ref}$ is the contribution of the variable component in the reference image, $f^{\rm ext}_{\rm sci}$ is the contribution from the extended component in the science image, and $f^{\rm var}_{\rm sci}$ is the contribution of the variable component in the science image. Note that the contribution of the extended component can vary between the reference and science images due to seeing effects, which can create an artificial source of variability. The scientifically relevant component for variability studies is the flux of the compact component, but it is difficult to separate it from the extended component. The second best alternative is to recover the flux of the compact component plus a constant contribution from the extended component. For this we can define an effective science flux, $\hat f_{\rm sci}$:
\begin{align}
    \hat f_{\rm sci} &\equiv f^{\rm ext}_{\rm ref} + f^{\rm var}_{\rm sci} \label{eq:fscicomp} \\
    &= f_{\rm ref} + {\rm sgn}~ f_{\rm diff} \label{eq:fsci},
\end{align}
which considers the same contribution of the extended component at all times. If the reference image changes, we can introduce a new effective science flux, $\hat f_{\rm ref, 0}$, that considers the contribution from the extended component from the first reference image used to generate alerts:
\begin{align}
    \hat f_{\rm sci, 0} &= f^{\rm ext}_{\rm ref, 0} + f^{\rm var}_{\rm sci} \\ 
    &= \hat f_{\rm sci} + \bigl(f^{\rm ext}_{\rm ref, 0} - f^{\rm ext}_{\rm ref})\label{eq:fsci_corr},
\end{align}
where $f^{\rm ext}_{\rm ref, 0}$ is the (unknown) contribution from the extended component from the first reference image. Note that the expected value from the second term is zero.

\subsection{Light Curve Variances}\label{eq:LCvar}

The computation of errors of the science flux must take into account the relation between the difference and reference fluxes, which are correlated. We can estimate the variance of the effective science flux, $\Var[\hat f_{\rm sci}]$, starting from Equation~\ref{eq:fsci} and using Equations~\ref{eq:frefcomp} and \ref{eq:sgnfdiff}:
\begin{align}
        \Var[\hat f_{\rm sci}] &= \Var[f_{\rm ref} + {\rm sign} ~ f_{\rm diff}] \\
        &= \Var[f_{\rm ref}] + \Var[f_{\rm diff}] + 2 ~\Cov[f_{\rm ref}, {\rm sign} ~ f_{\rm diff}] \\
        &= \Var[f_{\rm ref}] + \Var[f_{\rm diff}] + 2 ~\Cov[f^{\rm ext}_{\rm ref} + f^{\rm var}_{\rm ref}, f^{\rm var}_{\rm sci} - f^{\rm var}_{\rm ref}] \label{eq:varfsci_expansion}\\
        &= \Var[f_{\rm ref}] + \Var[f_{\rm diff}] - 2 ~\Var[f^{\rm var}_{\rm ref}]. \label{eq:varfsci_pre}
\end{align}
Note that the variance due to sky emission is contained in the first two terms of Equation~\ref{eq:varfsci}. One can also include additional terms in Equation~\ref{eq:varfsci_expansion} to reflect the contribution of the sky, but because these terms are not correlated they have no additional contribution in the covariance. We can expand  Equation~\ref{eq:varfsci_pre} to get the following:
\begin{align}
        \Var[\hat f_{\rm sci}] &= \Var[f_{\rm ref}] + \Var[f_{\rm diff}] - 2 ~\Var[f^{\rm var}_{\rm ref}] \\
        &= \Var[f^{\rm ext}_{\rm ref} + f^{\rm var}_{\rm ref}] + \Var[f_{\rm diff}] - 2 ~\Var[f^{\rm var}_{\rm ref}] \\
        &= \Var[f^{\rm ext}_{\rm ref}] + \Var[f^{\rm var}_{\rm ref}] + 2~\Cov[f^{\rm ext}_{\rm ref} , f^{\rm var}_{\rm ref}] + \Var[f_{\rm diff}] - 2 ~\Var[f^{\rm var}_{\rm ref}] \\
        &= \Var[f^{\rm ext}_{\rm ref}] + \Var[f^{\rm var}_{\rm ref}] + \Var[f_{\rm diff}] - 2 ~\Var[f^{\rm var}_{\rm ref}] \\
        &= \Var[f_{\rm diff}] - \Var[f^{\rm var}_{\rm ref}] + \Var[f^{\rm ext}_{\rm ref}]. \label{eq:varfsci}
\end{align}
and in the case of a change in the reference image, using Equation~\ref{eq:fsci_corr}, \ref{eq:varfsci} and \ref{eq:fscicomp}:
\begin{align}
        \Var[\hat f_{\rm sci, 0}] &= \Var[\hat f_{\rm sci} + (f^{\rm ext}_{\rm ref, 0} - f^{\rm ext}_{\rm ref})]\\
        &= \Var[\hat f_{\rm sci}] + \Var[f^{\rm ext}_{\rm ref, 0}] + \Var[f^{\rm ext}_{\rm ref}] - 2~\Cov[\hat f_{\rm sci}, f^{\rm ext}_{\rm ref}] \\
        &= \Var[f_{\rm diff}] - \Var[f^{\rm var}_{\rm ref}] + \Var[f^{\rm ext}_{\rm ref}] + \Var[f^{\rm ext}_{\rm ref, 0}] + \Var[f^{\rm ext}_{\rm ref}] - 2~\Cov[f^{\rm ext}_{\rm ref} + f^{\rm var}_{\rm sci}, f^{\rm ext}_{\rm ref}] \\
        &= \Var[f_{\rm diff}] - \Var[f^{\rm var}_{\rm ref}] + \Var[f^{\rm ext}_{\rm ref, 0}]. \label{eq:varfsci_corr}
\end{align}

To summarize, we show Equations~\ref{eq:fsci}, \ref{eq:fsci_corr}, \ref{eq:varfsci}, \ref{eq:varfsci_corr}:
\begin{align}
    \hat f_{\rm sci} &= f_{\rm ref} + {\rm sgn} f_{\rm diff} \notag \\
    \hat f_{\rm sci, 0} &= \hat f_{\rm sci} + (f^{\rm ext}_{\rm ref, 0} - f^{\rm ext}_{\rm ref}) \notag \\
    \Var[\hat f_{\rm sci}] &= \Var[f_{\rm diff}] - \Var[f^{\rm var}_{\rm ref}] + \Var[f^{\rm ext}_{\rm ref}] \notag \\
    \Var[\hat f_{\rm sci, 0}] &= \Var[f_{\rm diff}] - \Var[f^{\rm var}_{\rm ref}] + \Var[f^{\rm ext}_{\rm ref, 0}]. \notag
\end{align}
A problem with these formulae is that neither the variable nor extended components are known. However, they led us to consider the following cases:
\begin{enumerate}
    \item The contribution from the extended component is negligible in all the reference images:
    \begin{align}
        f^{\rm ext}_{\rm ref} &= 0 \notag \\
        &\Rightarrow \notag \\
        \hat f_{\rm sci, 0} = \hat f_{\rm sci} &= f_{\rm ref} + {\rm sgn} ~ f_{\rm diff} \\
        \Var[\hat f_{\rm sci, 0}] = \Var[\hat f_{\rm sci}] &= \Var[f_{\rm diff}] - \Var[f_{\rm ref}]. \label{eq:noext}
    \end{align}
    \item The contribution from the extended component is similar in all the reference images, and its contribution is similar to that from the variable component:
    \begin{align}
        f^{\rm ext}_{\rm ref,0} = f^{\rm ext}_{\rm ref}\ \ &\& \ \  f^{\rm var}_{\rm ref} = f^{\rm ext}_{\rm ref} \notag \\
        &\Rightarrow \notag \\
        \hat f_{\rm sci, 0} = \hat f_{\rm sci} &= f_{\rm ref} + {\rm sgn} ~ f_{\rm diff} \\
        \Var[\hat f_{\rm sci, 0}] = \Var[\hat f_{\rm sci}] &= \Var[f_{\rm diff}]. \label{eq:simext}
    \end{align}
    \item The contribution from the extended component is similar in all the reference images, and its contribution is dominant over the variable component:
    \begin{align}
        f^{\rm ext}_{\rm ref,0} = f^{\rm ext}_{\rm ref}\ \ &\& \ \  f^{\rm var}_{\rm ref} = 0 \notag \\
        \Rightarrow \notag \\
        \hat f_{\rm sci, 0} = \hat f_{\rm sci} &= f_{\rm ref} + {\rm sgn} ~ f_{\rm diff} \\
        \Var[\hat f_{\rm sci, 0}] = \Var[\hat f_{sci}] &= \Var[f_{\rm diff}] + \Var[f_{\rm ref}]. \label{eq:ext}
    \end{align}
\end{enumerate}
A visual inspection of variable star light curves confirms that Equation~\ref{eq:noext} is a better approximation in the case where there is no contribution from an extended component. In the case of AGNs, we have found that Equation~\ref{eq:simext} appears to be a better reflection of the measurement errors, which is consistent with having a similar contribution from the extended and variable components. In the case of transients, the extended component dominates the flux in the reference, but for these cases the scientifically relevant flux is the difference flux and its error. For this reason, we report the difference flux with its error, as well as the effective science flux with the errors (after a conversion of the fluxes to magnitudes) from Equations~\ref{eq:noext} and \ref{eq:simext} for every object where it is possible to correct the photometry, letting the users decide which flux and error to use for their particular science.

\subsection{Light curve magnitudes} 

The corrected photometry magnitude results from adding/subtracting the fluxes from the reference and difference in the same unit system and then converting to magnitudes. We can compute $\hat f_{\rm sci}$ by transforming the reference and difference magnitudes using the zero points of the science image:
\begin{align}
    \hat f_{\rm sci} &= f_{\rm ref} + {\rm sgn} ~ f_{\rm diff} = 10^{\frac{{\rm ZP}_{\rm sci} - m_{\rm ref}}{2.5}} + {\rm sgn} ~ 10^{\frac{{\rm ZP}_{\rm sci} - m_{\rm diff}}{2.5}}, \notag
\end{align}
where ${\rm ZP}_{\rm sci}$ is the zero point of the science image. This implies that the effective science magnitude, $\hat m_{\rm sci}$, will be:
\begin{align}
    \hat m_{\rm sci} &= -2.5 \log f_{\rm sci} + {\rm ZP_{\rm sci}} \notag \\
    &= -2.5 \log \bigl(10^{\frac{{\rm ZP}_{\rm sci} - m_{\rm ref}}{2.5}} + {\rm sgn} ~ 10^{\frac{{\rm ZP}_{\rm sci} - m_{\rm diff}}{2.5}}\bigr) + {\rm ZP}_{\rm sci} \notag \\
    &= -2.5 \log \bigl(10^{-\frac{m_{\rm ref}}{2.5}} + {\rm sgn} ~ 10^{-\frac{m_{\rm diff}}{2.5}}\bigr). \label{eq:msci}
\end{align}

Finally, we show the reported errors for Equations~\ref{eq:noext} and \ref{eq:simext}:
\begin{align}
    \delta \hat m_{\rm sci} &= \frac{\bigl(10^{-0.8~ m_{\rm diff}} \delta m_{\rm diff}^2  - 10^{-0.8~ m_{\rm ref}} \delta m_{\rm ref}^2\bigr)^{0.5}}
        {10^{-0.4~ m_{\rm ref}} + {\rm sgn}~ 10^{-0.4 ~m_{\rm diff}}},
\end{align}
to be used when there is no significant contribution from an extended component; or
\begin{align}
    \delta \hat m_{\rm sci} &= \frac{10^{-0.4~ m_{\rm diff}} \delta m_{\rm diff}}
        {10^{-0.4~ m_{\rm ref}} + {\rm sgn}~ 10^{-0.4 ~m_{\rm diff}}},
\end{align}
to be used when there is a contribution from an extended component, assumed to be similar to the variable component.

\setcounter{table}{0}
\renewcommand{\thetable}{B\arabic{table}}
\section{Tables}

Table~\ref{tab:etendue} provides the list of telescopes that was used in preparing Figure~\ref{fig:etendue}, along with their names and a relevant accompanying reference.

Tables~\ref{tab:MLvarI}, \ref{tab:MLvarII}, and \ref{tab:MLtransients} refer to a number of studies in which light curves were used to perform ML-based classification of variable and transient sources. Tables~\ref{tab:MLvarI} and \ref{tab:MLvarII} both refer to studies in which only persistent variable star classes were used; the former refers to papers published between 2017-2019, whereas the latter includes studies that appeared in print before 2017. Table~\ref{tab:MLtransients}, in turn, refers to those studies in which only transient sources were considered. These three tables have the same structure, with the reference given in the first column, an acronym for the source of the data given in the second column (with keys provided in Tables~\ref{tab:MLobssources} and \ref{tab:MLsyntheticsources} for empirical and synthetic data, respectively), the number of classes considered shown in the third column, and the fourth column displaying acronyms representing the actual classes that were considered in each case. These acronyms, along with the classes that they are intended to represent, are laid out in Tables~\ref{tab:variablesI} through \ref{tab:transients}.

In the case of Tables~\ref{tab:variablesI} and \ref{tab:variablesII}, the pulsating variable star classes are shown. Table~\ref{tab:variablesI} includes pulsating stars in the upper and lower main sequence, Cepheids, RR Lyrae, blue subdwarfs, and compact (WD) pulsators. Table~\ref{tab:variablesII}, in turn, includes red giant and supergiant pulsators.

Table~\ref{tab:variablesIII} presents a number of additional stellar variability classes, including eclipsing, eruptive, cataclysmic, and rotational variables. Additional classes that are shown in this table include microlensing events, R~CrB stars, Be stars, and X-ray binaries, among others. 

Primarily extragalactic variable sources are shown in Tables~\ref{tab:AGNs} and \ref{tab:transients}. In the case of \ref{tab:AGNs}, the variability is typically related to the presence of SMBHs, as in the case of AGNs and QSOs. Table~\ref{tab:transients}, in turn, includes primarily a variety of SN classes, although a few transient events of non-SN origin, such as TDEs and kilonovae, are also included. 

{\em We emphasize that the classes and associated taxonomies that are implied by Tables~\ref{tab:MLvarI} through \ref{tab:transients} do not reflect our own choices, but are rather simply a summary of what has been used in the ML literature to date.} In particular, the reader should be aware that the list of classes, as given, suffers from several shortcomings, such as being incomplete, containing redundant entries, and including classes that may not be sufficiently well defined. Still, our best effort to interpret what the different authors have intended to express in each case is reflected in these tables, with definitions given following, among others, the General Catalog of Variable Stars \citep[GCVS;][]{KholopovGCVS}, Variable Star Index \citep[VSX;][]{WatsonVSX}, and the broad overview of stellar variability classes presented in \cite{2015pust.book.....C}. In the future, as the ALeRCE project matures, we will work towards producing and refining our own taxonomy, which we will perfect along the way as we enter the LSST era.

\begin{deluxetable*}{ cccccc}
\tablecaption{Selection of telescopes shown in Figure~\ref{fig:etendue}.} \label{tab:etendue}
\tablehead{
\colhead{Short name} & \colhead{Long name}  & \colhead{Reference}
}
\startdata
ASAS-SN        & All--Sky Automated Survey for Supernova & \citet{2017PASP..129j4502K} \\
ATLAS         & Asteroid Terrestrial-impact Last Alert System  & \citet{2018PASP..130f4505T} \\
BlackGEM      & BlackGEM & \url{https://astro.ru.nl/blackgem/} \\
Blanco-DECam  & V\'ictor Blanco telescope -- Dark Energy Camera & \citet{2015AJ....150..150F} \\
Clay-MegaCam  & Clay Telescope -- Megacam & \citet{2015PASP..127..366M} \\
CFHT-MegaCam  & Canada France Hawaii Telescope -- Megacam & \citet{2003SPIE.4841...72B} \\
CRTS          & Catalina Real--Time Transient Survey (CSS, MLS, SSS) & \citet{2009ApJ...696..870D} \\
Euclid        & Euclid Mission & \citet{2011arXiv1110.3193L} \\
Evryscope     & Evryscope -- South & \cite{2015PASP..127..234L} \\
{\em Gaia}          & {\em Gaia} Mission & \citet{2018AA...616A...1G} \\
HATPI         & HATPI & \url{https://hatpi.org/science/}\\
{\em Kepler}        & {\em Kepler} Mission & \citet{2010Sci...327..977B} \\
KMTNet     & Korea Microlensing Transient Network & \citet{2016JKAS...49...37K} \\
KISO          & Kiso Observatory & \citet{2014PASJ...66..114M} \\
LS-QUEST      & La Silla 40$''$ ESO Schmidt Telescope -- QUEST camera & \citet{2004AJ....127.1158V} \\
LSST          & Vera C. Rubin Observatory Legacy Survey of Space and Time & \citet{2009arXiv0912.0201L} \\
PanSTARRS     & Panoramic Survey Telescope and Rapid Response Response System & \citet{2002SPIE.4836..154K} \\
PTF           & Palomar Transient Factory & \citet{2009PASP..121.1395L} \\
SDSS          & Sloan Digital Sky Survey & \citet{2000AJ....120.1579Y} \\
Subaru-HSC    & Subaru telescope  -- Hyper Suprime-Cam & \citet{2018PASJ...70S...4A} \\
SkyMapper     & SkyMapper Southern Sky Survey & \citet{2007PASA...24....1K} \\
TESS          & Transiting Exoplanet Survey Satellite & \citet{2015JATIS...1a4003R} \\
VISTA         & Visible and Infrared Survey Telescope for Astronomy & \citet{2006SPIE.6269E..0XD} \\
VST-OmegaCam  & VLT Survey Telescope -- OmegaCam & \citet{2005Msngr.120...13C} \\
WFIRST        & Wide Field Infrared Survey Telescope & \citet{2015arXiv150303757S} \\
        &  (aka Nancy Grace Roman Space Telescope) & \\
ZTF           & Zwicky Transient Facility & \citet{2019PASP..131a8002B}
\enddata
\end{deluxetable*}

\begin{deluxetable*}{cccc}
\tablecaption{Light curve based ML classifiers that include only persistent variable classes (more than 2 classes) between 2017 and 2019. Class abbreviations are defined in Tables~\ref{tab:variablesI} to \ref{tab:transients}} \label{tab:MLvarI}
\tablehead{
\colhead{\small{Reference}}  & \colhead{\small{Data source}} & \colhead{\small{\#classes}} & \colhead{\normalsize{classes}} \\
}
\startdata
\cite{2019AA...625A..97R} & {\em Gaia} DR2 & 18 & E,~CV,~RSCvn,~BLAP,\\
                            & & &  Mira+SR,~DSCT+SXPh,~RRL(ab,~c,~d,~Ad),\\
                            & & & CephCl,~ACEP,~CephII,\\
                            & & & Low amp.:DSCT+GDOR,~ELL,~OSARG,~FL+ROT,~Other \\
\cite{2019ApJ...877L..14T} & ASAS-SN & 8 & DSCT,~RRL(ab,~cd),~Ceph,~E,~ROT,\\
                            & & & Mira,~SR \\
\cite{2019MNRAS.486.1907J} & ASAS-SN & 10 & Ceph,~DSCT,~E(EW,EA|EB,EB),~RRL(ab,c),\\
                            & & & M,~SR,~Irregular\\
\cite{2019MNRAS.488.4858H} & CSDR2 & 12 & RRL(ab,~c,~d),~Blazhko,~E(C+SD,D),\\
                            & & & ROT,~LPV,~DSCT,~Ceph(II,A) \\
\cite{2019MNRAS.tmp.2752J} & UCR & 3 & RRL,~Ceph,~E \\
                            & LINEAR & 5 & RRL(ab,~c),~DSCT,~E(C,SD) \\
\cite{2019MNRAS.482.5078A} & OGLE+VVV & 9 & Ceph(F,~01),~RRL(ab,~c),\\
                            & +CoRoT & & E(C,~SD+D),~Mira,~SR,~OSARG\\
\cite{2018AJ....155...16C} & MACHO & 8 & NV,~QSO,~BeS,~Ceph,~RRL,~E,~ML,~LPV \\ 
                            & OGLE & 6 & Ceph,~CephII,~RRL,~E,~DSCT,~LPV \\
\cite{2018NatAs...2..151N} & ASAS & 5 & RRLab,~Ceph,~SR,~BPer,~WUMa \\
                            & LINEAR & 5 & DSCT,~RRL(ab,~c),~BPer,~WUMa \\
                            & MACHO & 8 & Ceph(F,~O1),~LPVW,~RRL(ab,~c,~e,~GB)\\
\cite{2018MNRAS.474.3259V} & OGLE & 8 & Ceph(CL,~II,~A),~RRL,~LPV,~DPV,~DSCT,~E\\
 							& MACHO & 11 & RRL(ab,~c,~e,~GB),~Ceph(F,~O1),\\
 							& & & LPVW(A,~B,~C,~D),~E\\
\cite{2017arXiv170906257M} & CSDR2 & 7 & E(C,~SD),~RRL(ab,~c,~d),~RSCVn,~LPV \\
\cite{2017ApJ...845..147B} & EROS, & 5 & Ceph,~E,~QSO,~RRL,~LPV\\
 & MACHO,~HiTS & &  \\
\cite{2017MNRAS.468.2189Z} & OGLE & 8 & Mira,~QSO,~SR,~OSARG,~Ceph(F,~O1),\\ & & & RRL(ab+d,~c+e)\\
\enddata
\end{deluxetable*}

 \begin{deluxetable*}{ cccc}
\tablecaption{Light curve based ML classifiers that include only persistent variable objects (more than 2 classes) before 2017. Class abbreviations are defined in Tables~\ref{tab:variablesI} to \ref{tab:transients}} \label{tab:MLvarII}
\tablehead{
\colhead{\small{Reference}} & \colhead{\small{Data source}} & \colhead{\small{\#classes}} & \colhead{\normalsize{classes}} \\
}
\startdata
 \cite{2016AA...587A..18K} & MACHO, & 19 & DSCT,~RRL(ab,~c,~d,~e),\\
 & LINEAR,~ASAS & & Ceph(F,~O1,~other,~II),~E(C,~SD,~D),\\
                &  & & LPV(MAGBC,~MAGBO,~OSARGAGB,\\
                & & &  OSARGRGB,~SRAGBC,~SRAGBO),~NV\\
 \cite{2016ApJ...820..138M} & OGLE & 6 & Ceph(CL,~II),~RRL,~E,~DSCT,~LPV \\
 							& MACHO & 8 & NV,~QSO,~BeS,~Ceph,~RRL,~E,~ML,~LPV \\
 \cite{2016ApJ...819...18P} &  MACHO & 8 & BeS,~Ceph,~E,~LPV,~ML,~NV,~QSO,~RRL\\
 							& EROS & 11 & E,~RRL,~Ceph(F,~O1,~DM,~II),\\
 							& & & LPV(OSARGRGBO,~SRAGBO,\\
 							& & & SRAGBC,~MAGBC,~MAGBO)\\
 \cite{2016AJ....152...71N} & MACHO & 8 & NV,~QSO,~BeS,~Ceph,~RRL,~E,~ML,~LPV\\
 \cite{2016MNRAS.459.3721B} & {\em Kepler} & 14 & ACT,~BCep,~Ceph,~DSCT,~E,~ELL,~GDor,~ROT,\\
 							& & & RRL(ab,~c),~RVTau,~SPB,~SR,~MISC/NV \\
 \cite{2016SADM....9....1F} & & & \\
\cite{2015MNRAS.451.3385K} & OGLE & 3 & Ceph,~E,~RRL \\
							&  ASAS & 7 &  Mira,~RRLab,~E(C,~D,~SD),~DSCT,~CephF \\
\cite{kim2014epoch} &  EROS-2 & 26 & DSCT,~RRL(ab,~c,~d,~e),~Ceph(F,~O1,~Other),~CephII\\
                            & & & E(C,~SD,~D,~SD+D,~Other),~BeS,~QSO,~NV\\
                            & & & LPV(MAGB(C,~O),~OSARGAGB(C,~O),\\
                            & & & OSARGRGB(C,~O),~SRAGB(C,~O))\\
\cite{2013ApJ...777...83P} & SAGE, 2MASS, & 7 & NV,~QSO,~BeS,~Ceph,~RRL,~E,~LPV\\
 							& UBVI, MACHO & &  \\
 \cite{2012ApJS..203...32R} & ASAS  & 28 & DSCT,~SXPh,~RRL(ab,~c,~d),~Ceph(CL,~MM,~II),\\
 							& & & Mira,~SR,~LPVW(A,~B),~RVTau,~BCep,~RSG, \\
                            & & & BPer,~BLyr,~WUMa,~ChemPec,~ELL,~RSCvn, \\
                           	& & & HAeBe,~CTTau,~WLTTau,~RCB,~LBV,~BeS\\
\cite{2009AA...506..519D} & CoRoT & 29 & sdBV,~DSCT,~LBoo,~SXPh,~roAp,~GDor,\\
                            & & & RR(ab,~c,~d),~Ceph(CL,~DM,~II),~RVTau,\\
                            & & & Mira,~SR,~PVSG,~BCep,~SPB,~E,\\
                            & & & ChemPec,~ELL,~FUOri,~HAeBe,~TTau,\\
                            & & & LBV,~WR,~XB,~BeS,~LAPV\\
\cite{2007AA...475.1159D}  & OGLE & 35 & DAV,~DBV,~sdBV,~GWVir,\\
                            & & & DSCT,~LBoo,~SXPh,~roAp,~GDor,\\
                            & & & RRL(ab,~c,~d),~Ceph(Cl,~DM,~II),\\
                            & & & PVSG,~Mira,~SR,~RVTau,~BCep,~SPB,\\
                            & & & E(C,~SD,~D),~ChemPec,~ELL,\\
                            & & & FUOri,~HAeBe,~TTau,~LBV, \\
                            & & & SLR,~WR,~XB,~CV,~BeS
\enddata
\end{deluxetable*}

\begin{table*} 
\centering
\caption{Light curve based ML classifiers that include only transient objects. Class abbreviations are defined in Table~\ref{tab:transients}} \label{tab:MLvarIII} \label{tab:MLtransients}
\vspace{.2cm}
\begin{tabular}{ cccc}
Reference  & Data source & \#classes & \normalsize{classes} \vspace{.2cm}  \\
\hline
\cite{2019ApJ...884...83V} & PS1-MDS & 5 & SNIa,~SNIbc,~SNII,~SNIIn,~SLSN \\
\cite{2019PASP..131k8002M} & PLAsTiCC & 12 & TDE,~CART,~ILOT,~PISN,~kN,~.Ia,\\
                            & & & SNIa,~SNIax,~SNIa-91bg,~SNIbc,~SNII \\
\cite{2019MNRAS.tmp.2989M} & SNANA & 2 & SNIa,~other \\
\cite{2019arXiv190100461B}  & SNANA, SPCC & 2 & SNIa,~other\\
 \cite{2018MNRAS.473.3969R} & SPCC & 3 & SNIa,~SNII,~SNIbc \\
\cite{2017ApJ...837L..28C} & SPCC & 3 & SNIa,~SNII,~SNIbc \\
 \cite{2016ApJS..225...31L} & SPCC & 3 & SNIa,~SNII,~SNIbc \\
 \cite{2013MNRAS.429.1278K} & SPCC & 2 & SNIa,~other\\
 \hline
\end{tabular}
\end{table*}

\begin{table*}[h!] 
\centering
\caption{Observational data sources used for ML classification.} \label{tab:MLobssources}
\vspace{.2cm}
\begin{tabular}{ ccc}
Abbreviation  & Long name & Reference \vspace{.2cm}  \\
\hline
ZTF & Zwicky Transient Facility &  \cite{2019PASP..131a8002B} \\
HSC--SSP & Hyper Suprime-Cam Subaru Strategic Program & \cite{2018PASJ...70S...4A} \\
UCR & University of California Riverside & \cite{UCRArchive2018}\\
    & Time Series Classification Archive \\
OSC & Open Supernova Catalog & \cite{2017ApJ...835...64G} \\
ASAS-SN & All-Sky Automated Survey for Supernovae & \cite{2017PASP..129j4502K}\\
CSDR2 & The Catalina Surveys Data Release 2 &  \cite{2017MNRAS.469.3688D}\\
HiTS & High cadence Transient Survey & \cite{2016ApJ...832..155F} \\
PS1-MDS & Pan-STARRS-1 Medium Deep Survey & \cite{2011AAS...21832812H}\\
LINEAR & Lincoln Near-Earth Asteroid Research Survey & \cite{2011AJ....142..190S} \\
UBVI  & UBVI photometry of six open cluster candidates  & \cite{2011NewA...16..161P} \\
VVV & Vista Variables in the Via Lactea & \cite{2010NewA...15..433M} \\
OGLE & The Optical Gravitational Lensing Experiment & \cite{udalski2008optical}\\
2MASS & The Two Micron All Sky Survey &  \cite{2006AJ....131.1163S} \\
SAGE & Spitzer Survey of the Large Magellanic Cloud: & \cite{2006AJ....132.2268M} \\
 & Surveying the Agents of a Galaxy's Evolution & \\
CoRoT & Convection, Rotation, and planetary Transits & \cite{2006cosp...36.3749B} \\
SDSS & The Sloan Digital Sky Survey & \cite{2000AJ....120.1579Y} \\
MACHO & Massive Compact Halo Objects survey & \cite{2000ApJ...542..281A} \\
EROS & Exp\'erience pour la Recherche d'Objets Sombres & \cite{1998AA...332....1P}\\
ASAS & All Sky Automated Survey & \cite{1997AcA....47..467P} \\
\hline
\end{tabular}
\end{table*}

\begin{table*}[h!] 
\centering
\caption{Synthetic data sources used for ML classification.}
\label{tab:MLsyntheticsources}
\vspace{.2cm}
\begin{tabular}{ ccc }
Abbreviation  & Long name / Description & Reference \vspace{.2cm}  \\
\hline
PLAsTiCC & Photometric LSST  Astronomical & \cite{2019PASP..131i4501K} \\
        & Time-Series Classification Challenge &  \\
SNANA & SuperNova ANAlysis software & \cite{2009PASP..121.1028K}\\
SPCC & Supernova Photometric Classification Challenge & \cite{2010arXiv1001.5210K} \\
 & Type II SNe confined wind acceleration model & \cite{2019MNRAS.488.3949M} \\
 & Type Ia SNe spectral templates & \cite{2007ApJ...663.1187H}\\
\hline
\end{tabular}
\end{table*}

\begin{table*}
\centering
\caption{Pulsating variable star classes (excluding red giants and supergiants) found in the ML literature (see text for further details).}
\label{tab:variablesI}
\begin{tabular}{>{\footnotesize}p{1.5cm} | >{\footnotesize}p{2.8cm}>{\footnotesize}p{11.5cm}}
\small{Type } & \small{Class abbrev.}  & \small{Brief description} \vspace{.2cm} \\
\hline
\multirow{9}{*}{Lower MS}
& DSCT &  $\delta$ Scutis. Low-order $p$-mode pulsators. Both radial and non-radial modes can be present. Periods typically shorter than 0.42~d. Pop.~I. \\
& LBoo & $\lambda$ B\"ootis. A--type MS dwarf with low metallicities. Part of the DSCT class. \\
& SXPh & SX Phoenicis. Pop.~II counterparts of the DSCT. Typically found in globular clusters and dSph galaxies. Includes pulsating blue straggler stars. \\
& roAp & Rapidly oscilating Ap stars. High-order, non-radial $p$-mode pulsators. Amplitudes typically do not exceed 0.012~mag in $V$. \\
& GDor & $\gamma$ Doradus. High-order, non-radial $g$-mode pulsators. Periods between 0.3 and 3~d, amplitudes less than 0.1~mag in $V$. \\
\hline
\multirow{3}{*}{Upper MS}
& BCep & $\beta$ Cepheids. Non-radial $p$-mode pulsators. Periods between 0.1--0.6~d, amplitudes in $V$ between 0.01--0.32~mag. \\
& SPB & Slowly pulsating blue stars, aka 53~Per stars. Non-radial $g$-mode pulsators. Periods between 0.4--6~d, amplitudes in $V$ less than 0.03~mag.  \\
\hline
\multirow{3}{*}{RR Lyrae}
& RRL(\footnotesize{ab,c,d,Ad,e,GB}) & RR Lyrae. Pulsating horizontal-branch stars, with periods of order 0.5~d. Subtypes: ab (fundamental-mode), c (first overtone), d (double-mode), Ad (anomalous double mode), e (second overtone). Also classified by location (Galactic bulge, GB). \\
& Blazhko & RRL with long-period modulations (Blazhko effect). \\
\hline
\multirow{8}{*}{Cepheids}
& Ceph(\footnotesize{CL,F,O1, DM,MM,other})  & $\delta$ Cepheids, aka classical (CL) Cepheids or type~I Cepheids. Pulsating G-K giant and supergiant stars. Often found pulsating in the fundamental (F), first (OI), or second overtone; double (DM) or multi-mode (MM) pulsation also common. \\
& ACEP & Anomalous Cepheids, aka BL~Boo stars. Evolved counterparts of the SX~Phe stars. Commonly found in dSph galaxies. \\
& CephII & Type II Cepheids. Low-mass Pop. II stars, often subdivided into BL~Her, W~Vir, and RV~Tau subclasses with increasing periods. \\
& RVTau & Type II Cepheids with periods in excess of 30~d. Light curves are well-behaved and show double minima at the short-period end, but become increasingly irregular with increasing period. \\
\hline
\multirow{3}{*}{Subdwarf}
& sdBV & Pulsating subdwarf B stars, aka V361~Hya, EC~14026, sdBV$_p$, or sdBV$_r$ stars. $p$-mode pulsators in which both radial and non-radial modes can be present. Periods between 60 and 570~s, amplitudes in $V$ less than 65~mmag. \\
\hline
\multirow{5}{*}{Compact }
& GW Vir & Pulsating pre-WD stars, aka pulsating PG~1159 stars. Includes both pulsating O-type WD stars (DOVs) and so-called planetary nebulae nucleus variables (PNNVs). \\
& DAV & Pulsating A-type WD star, aka ZZ~Ceti variables. Non-radial $g$-mode pulsators with H-dominated atmospheres. \\
& DBV & Pulsating B-type WD stars, aka V777~Her stars. Non-radial $g$-mode pulsators with He-dominated atmospheres. \\
\hline
\end{tabular}
\end{table*}

\begin{table*}
\centering
\caption{As in Table~\ref{tab:variablesI}, but for pulsating red giants and supergiants.}
\label{tab:variablesII}
\begin{tabular}{>{\footnotesize}p{1.5cm} | >{\footnotesize}p{2.8cm}>{\footnotesize}p{11.5cm}}
\small{Type} & \small{Class abbrev.}  & \small{Brief description} \\
\hline
\multirow{13}{*}{Red Giants }
& LPV & Long Period Variable. Pulsating cool giant or supergiant stars. Often subdivided into Miras, SRs, Irregulars, and OSARGs. \\
& Mira & Mira variables. LPV red giants with very red colors and large amplitudes (by definition, exceeding 2.5~mag in $V$). Can be C- or O-rich, depending on evolutionary history. \\
& SR & Semi-regular variables. Similar to the Miras, but with smaller amplitudes (by definition, not exceeding 2.5~mag in $V$). Often subdivided into SRa (persistent periodicity), SRb (poorly defined periodicity), SRc (red supergiant SRs), and SRd (orange/yellow supergiant SRs). \\
& OSARG & OGLE Small Amplitude Red Giant. Less evolved/luminous counterpart of the Miras and SRs, with smaller amplitudes and frequently multiple pulsation modes present. \\
& LPVW(\footnotesize{A,B,C,D}) & LPVs classified according to the sequence that they follow in a so-called Wood diagram \citep{pwea1999}. \\
& LPV(\footnotesize{MAGB[C,O]}) & C- or O--rich  Mira-type LPVs on the asymptotic giant branch (AGB) \\
& LPV(\footnotesize{OSARGAGB}) & OSARG-type LPVs on the AGB \\
& LPV(\footnotesize{OSARGRGB[O]}) & Normal or O--rich OSARG-type LPVs on the red giant branch \\
& LPV(\footnotesize{SRAGB[C,O]}) & C- or O--rich SR-type LPVs on the AGB \\
\hline
\multirow{5}{*}{Supergiants} & RSG & Red supergiant stars with irregular or semi-regular light curves (Lc and SRc, respectively, as per the GCVS). According to \citep{chatys2019},  periodicities may include two groups, related to pulsations ($P \sim 300-1000$~d) and LSPs ($P \sim 1000-8000$~d). \\
& LSP & LPV red giants with long secondary periods. \\
& PVSG & Periodic variable supergiant star.\\
\hline
\end{tabular}
\end{table*}

\begin{table*}[h!] 
\centering
\caption{Stellar variability classes,  other than the pulsating ones, in the ML literature  (see text for further details).} 
\label{tab:variablesIII}
\vspace{.2cm}
\begin{tabular}{>{\footnotesize}p{1.5cm} | >{\footnotesize}p{2.8cm}>{\footnotesize}p{11.5cm}}
\small{Var. Type} & \small{Class}  & \small{Brief description} \vspace{.2cm} \\
\hline
Non--variable & NV & Non--variable star \\
 \hline
\multirow{6}{*}{Eclipsing}
& E(C,~SD,~D) & Eclipsing binary, classified according to its physical status as contact (C), semi-detached (SD), or detached (D)\\
& BPer, BLyr, WUMa & Eclipsing binary, phenomenologically classified, according to its light curve shape, into $\beta$~Per (Algol, EA), $\beta$~Lyr (EB), and W~UMa (EW), respectively. \\ 
\hline
\multirow{6}{*}{Rotational} & ROT & Rotational variable. Rotating stars with non--uniform surface (starspots).  \\
 & ChemPec & Chemically peculiar rotational variable star.\\
 & ELL & Close binary systems with ellipsoidal components (not eclipsing). \\
 & RSCVn & RS Canum Venaticorum variable. Binary systems in which the primary star is typically a giant, characterized by semi--periodic light curves due to active chromospheres and the presence of starspots. \\
\hline
\multirow{2}{*}{Chromosph.} & ACT & Stars presenting surface  activity due to active coronae and chromospheres. \\
& Mdwarf & M--dwarf flaring star; flares are caused by magnetic field reconnection events. \\
\hline
 & [C,WL]TTau & Classic (C) or weak-lined (WL) T~Tauri stars. Low-mass YSOs undergoing accretion from their surrounding disks. Depending on the H$\alpha$ emission strength, they are subdivied into C (strong emission) and WL (weak emission). Possible evolutionary link with EX Lupi (EXor) and FU Ori (FUor) stars, according to the mass accretion rate. \\
\multirow{5}{*}{YSO} & HAeBe & Herbig Ae/Be star. 
 Higher-mass counterparts of the T~Tauri stars. When large, irregular dust obscuration events are present, they may also be classified as UX~Ori (UXor) stars. \\
 & FUOri & FU Orionis stars. Pre--MS stars undergoing abrupt mass accretion episodes.\\
\hline
\multirow{3}{*}{Outburst} & LBV & Luminous blue variable (aka S Doradus) star. Hot, luminous stars near or above the Eddington limit undergoing vigorous mass loss and outbursts, followed by quiescent states.  \\
 & CV/Nova & Cataclysmic variable star (including classical novae). Mass transferring binary system in which a MS star transfers mass onto a WD via Roche Lobe overflow. In the case of classical novae, thermonuclear explosions take place at the surface of the mass-accreting WD, followed by a quiescent state. \\
\hline
Lensing & ML & Microlensing event. Star whose brightness is magnified due to a gravitational lensing event. \\
\hline
\multirow{8}{*}{Other} & RCB & R Coronae Borealis stars. F- or G-type self-eclipsing supergiant stars that undergo dramatic dimming events, brought about by mass loss episodes followed by dust condensation. \\
 & DPV & Double periodic variable. Binary system with variability due to eclipses or ellipsoidal modulations on timescales of order a few days, accompanied by a long cycle lasting about 33 times the orbital period.\\
 & BeS & Be stars. Non--supergiant B star rotating close to break-up speed and presenting decretion disks, accompanied by variable Balmer emission. \\
 & LAPV & Low amplitude periodic variable. Defined in \cite{2009AA...506..519D}, include low-amplitude Cepheids and also rotational variable stars with regular light curves. \\
 & WR & Wolf--Rayet star. Evolved, massive stars that have lost their H envelopes and show signatures of strong stellar winds.\\
 & XB & X--ray binary. CV-like systems in which the accreting star is typically not a WD, but rather a neutron star or black hole, and which thus emit their energy mostly in the form of X rays. \\
 \hline
\end{tabular}
\end{table*}

\begin{table*}[h!]
\centering
\caption{Extragalactic BH-related variability classes, as found in the ML literature (see text for further details).}
 \label{tab:AGNs}
\vspace{.2cm}
\begin{tabular}{>{\footnotesize}p{1.5cm} >{\footnotesize}p{11.5cm}}
Abbreviation  & Description \vspace{.2cm} \\
\hline
AGN & Active Galactic Nuclei. Central accreting SMBH (${>}10^{5}$\,$M_{\odot}$) where the host galaxy dominates the total light. Variability likely due to accretion-disk instabilities. \\
QSO & Quasi Stellar Object. Central accreting SMBH which  dominates over the host galaxy in the total light. Variability likely due to accretion-disk instabilities. \\
Blazar & Central accreting SMBH with a relativistic jet directed towards the observer. Variability due to sychrotron and inverse-compton relativistic beaming. This category does not distinguish between Blazars, 
BL Lacs, and optical violent variables (OVVs), which peak in different wavebands. \\
\hline
\end{tabular}
\end{table*}

\begin{table*}[h!] 
\centering
\caption{Transient classes, as found in the ML literature (see text for further details).}
\label{tab:transients}
\vspace{.2cm}
\begin{tabular}{ c >{\footnotesize}p{12cm}}
Abbreviation  & Description \vspace{.2cm} \\
\hline
SNIa & Type Ia supernova (SN). Thermonuclear explosion of a CO white dwarf. \\
SNIa-91bg & Underluminous SNe Ia. SN1991bg--like. \\
SNIax & Type Iax SNe. Deflagration dominated SN Ia. \\
.Ia & ``.Ia" SNe. He shell detonation explosion. \\
SNIbc & Type Ib or Ic SNe. Core collapse (CC) of envelope--stripped massive star.\\
SNII & Type II SNe. CC of red supergiant star. \\
SNIIn & Type IIn SNe. SN explosion in dense circumstellar medium. \\ 
TDE & Tidal Disruption Event. Stellar disruption due to BH proximity. \\
CART & Calcium Rich Transient. \\
ILOT & Intermediate Luminosity Optical Transient. \\
PISN & Pair instability SNe. CC and thermonuclear explosion due to e$^-$/e$^+$ pair production. \\
SLSN & Super Luminous SNe. Class of explosions about 10 times brighter than standard SNe.\\
kN & Kilonova. Neutron star merger optical counterpart. \\
\hline
\end{tabular}
\end{table*}


\bibliography{ALeRCE_WP}{}
\bibliographystyle{aasjournal}



\end{document}